\documentclass[11pt,a4paper]{article}
\usepackage{jheppub}

\newcommand{\del}{\partial}
\newcommand{\delbar}{\bar\del}
\newcommand{\tr}{\text{tr}}

\title{The FZZ duality with boundary}

\arxivnumber{1012.4731}

\author[a]{Thomas Creutzig,}
\author[b]{Yasuaki Hikida}
\author[c]{and Peter B. R\o nne}

\affiliation[a]{Department of Physics and Astronomy, University of North Carolina,\\
Phillips Hall, CB 3255, Chapel Hill, NC 27599-3255, USA}
\affiliation[b]{Department of Physics, and Research and Education
Center for Natural Sciences,\\
Keio University, Hiyoshi, Yokohama 223-8521, Japan}
\affiliation[c]{National Institute for Theoretical Physics and Centre
for Theoretical Physics,\\ University of the Witwatersrand, Wits, 2050, South Africa}

\emailAdd{creutzig@physics.unc.edu}
\emailAdd{hikida@phys-h.keio.ac.jp}
\emailAdd{peter.roenne@wits.ac.za}

\abstract{The Fateev-Zamolodchikov-Zamolodchikov (FZZ) duality relates Witten's cigar model to sine-Liouville theory. This duality was proven
in the path integral formulation and extended to the case of higher genus closed
Riemann surfaces by Schomerus and one of the authors.
In this note we further extend the duality
to the case with boundary. Specifically, we relate D1-branes in the cigar
model to D2-branes in the sine-Liouville theory. In particular, the boundary
action for D2-branes in the sine-Liouville theory is constructed.
We also consider the fermionic version of the FZZ duality.
This duality was proven as a mirror symmetry by Hori and Kapustin, but we
give an alternative proof in the path integral formulation which directly relates correlation functions.
Also here the case with boundary is investigated and the results are
consistent with those for branes in ${\cal N}=2$ super
Liouville field theory obtained by Hosomichi.}

\keywords{Conformal Field Models in String Theory, Boundary Quantum Field Theory, String Duality, Supersymmetry and Duality}

\begin{document}

\maketitle
\flushbottom

\section{Introduction}
\label{Intoduction}

In this paper we derive the Fateev-Zamolodchikov-Zamolodchikov (FZZ) duality
\cite{FZZ2} in the case where the world-sheet has a boundary.
The duality is between Witten's sigma model
\cite{Witten:1991yr} for the Euclidean 2d black hole and sine-Liouville field theory.
We also consider the fermionic version of the black hole sigma model and its duality with the $\mathcal N=(2,2)$ super-Liouville field theory.
The fermionic FZZ duality was proven as a mirror symmetry in \cite{HK}.
The bosonic two-dimensional black hole is described by the $H_3^+/\mathbb R$ coset sigma model with world-sheet coupling
given by the level $k$. Here $H_3^+=\mathrm{SL(}2,\mathbb C)/\mathrm{SU(}2,\mathbb R)$ is the Euclidean version of AdS$_3$.
The geometry describes a semi-infinite cigar with asymptotic radius $\sqrt{k}$.
In this paper we work in units where the string length is set to
$l_s \equiv \sqrt{\alpha '} = 1$.
The dual sine-Liouville theory has interaction terms which are defined on a cylinder with the inverse radius $1/\sqrt{k}$. The potential gives exponential suppression in one direction of the cylinder, but with a coupling constant which is the inverse as on the cigar side.
For large $k$ the cigar model is at weak curvature and thus has weak world-sheet coupling, while the dual sine-Liouville
potentials give strong coupling, and vice versa.

The main motivation for studying these target space dualities is their strong/weak coupling nature. Such dualities allow us to do calculations in the strongly coupled regions which are normally very hard to access. Calculations at strong coupling are important in many settings with a prominent example being the AdS/CFT duality \cite{Maldacena:1997re}. Here the string side with strong world-sheet coupling should be compared with a weakly coupled gauge theory dual, and being able to calculate on both sides will allow non-trivial checks of the conjecture. The FZZ duality involves Euclidean AdS$_3$ and has been used to get quantum corrections to the thermodynamics of the two-dimensional black hole via holography \cite{KKK}.
The fermionic version would be useful to study holography involving
NS5-branes as, e.g., in \cite{Giveon:1999px}.

The bosonic FZZ duality on closed world-sheets has recently been proven by Schomerus and one of the authors \cite{HS3} using a path-integral method.
Let us remind ourselves of the key steps in the proof.
First the coset $H_3^+/\mathbb R$ is written as a gauged WZNW model and embedded in the $H_3^+\times U(1)$
model \cite{Gawedzki:1991yu}. One then uses the fact
that the correlators of $H_3^+$ model can be written in terms of Liouville field
theory correlators with extra insertions of degenerate fields \cite{RT,HS}. The essential non-perturbative step is here to use Liouville theory's strong/weak self-duality.
Then, some treatments of fields and operators
 yield the correlators of
sine-Liouville theory.
The supersymmetric version of the FZZ duality has been derived by Hori and Kapustin \cite{HK} as mirror duality using the
standard, but rather indirect method of gauged linear sigma models.
In this paper we give an alternative proof using the method of \cite{HS3} and thus establish a direct path-integral derivation of mirror duality.
This also has the advantage that we do not only get a duality for the actions, but we
have a precise relation of correlation functions including the coupling dependent normalization.
Further, it extends readily to higher genus closed world-sheets \cite{HS3}.

The main part of this note is devoted to extend the FZZ and mirror duality to the case with open strings.
It is certainly an important problem to understand boundary conformal field theories with non-compact target spaces.
In eventually solving these models, knowledge of the boundary Lagrangian complying with the boundary conditions is important.%
\footnote{Examples of such models studied recently are deformations of Liouville field theory and $H_3^+$ model \cite{Babaro:2010em},
as well as supergroup WZNW models \cite{Creutzig:2010zp,Creutzig:2009zz,Creutzig:2008an,Creutzig:2008ag,Creutzig:2008ek,Creutzig:2007jy}.
Finding boundary actions is in general a difficult task and has not been solved in generality, see however \cite{Creutzig:2010ne}.}
In our case such a Lagrangian description is essential since we need a path integral formulation of correlation functions
to derive the dualities. Such a boundary action can luckily be found for AdS$_2$-type branes in the $H_3^+$ model \cite{FR}.
In our coset $H_3^+/\mathbb R$ such branes descend to 1-dimensional branes on the cigar.
Using the boundary action we can then derive the duality in a similar manner as the bulk case.
As with T-duality we expect Dirichlet and Neumann boundary conditions to be exchanged in the circular direction.
Indeed, the corresponding dual branes are of the two-dimensional FZZT-type \cite{FZZ,Teschner:2000md}.
As a result correlators of sine-Liouville and the cigar coincide, and thus the known disc one-point functions of the cigar \cite{RS} can be used to write down those of sine-Liouville. Moreover,
we obtain a Lagrangian for the boundary sine-Liouville theory which also allows to compute correlators directly in this model.

The article is organized as follows: In section two we consider the bosonic FZZ duality for open strings. In section 2.1 we give the action for the disk including the boundary part when we consider D1-branes.\footnote{Since we are in Euclidean space, D$p$-branes denote $p$-dimensional branes.} In section 2.2 the vertex operators are considered and the correlation functions are written in the path integral formalism. The correspondence to Liouville field theory is shown in section 2.3, and using the Liouville self-duality, the duality with sine-Liouville theory is derived in section 2.4. In section 3 we consider the supersymmetric FZZ duality, starting with the path integral formulation in section 3.1. Finally, the duality is derived for respectively closed and open world-sheets in sections 3.2 and 3.3. Conclusions are given in section 4. The paper is closed with five appendices. In appendix A we recall the geometry of the branes that we consider, and their possible boundary actions. The gauged sigma model describing the cigar is considered in appendix \ref{app:gaugedsigma} where we also explain the Chan-Paton factors appearing in our calculations. Some Jacobians from the change of measure are calculated in appendix \ref{Sklyanin}. In appendix \ref{ra} we derive the reflection coefficient of vertex operators that we need in the proofs of the dualities.
Finally, in appendix \ref{SLbranes} we discuss branes in sine-Liouville
theory.

\section{Boundary FZZ duality}

The FZZ duality conjecture was proven in \cite{HS3} and generalized to
correlation functions on closed Riemann surfaces of arbitrary genus.
In this section we would like to extend the duality to disk amplitudes.
Before proceeding let us briefly state the main result.
For the derivation of the FZZ duality with boundary, the main ingredient is to show the identity between correlators in the cigar model and sine-Liouville theory on the disk or equivalently upper half-plane
\begin{align}\label{eq:finalresultbosonic}
 &\left \langle \prod^N_{a = 1} \Psi^{j_a}_{m_a , \bar m_a} (z_a)
  \prod^M_{c = 1} \Psi^{l_c, i_c}_{m_c } (u_c)
 \right \rangle  ^\text{cig}
  \\ & = {\cal N}
  \left \langle \prod^N_{a = 1} e^{ 2 b (j_a+1) \phi + i \frac{2}{\sqrt{k}}
 ( m_a X_L -  \bar m_a  X_R )} (z_a)
  \prod^M_{c = 1} \frac{\sigma_{i_c}}{\sqrt{2}}
e^{  b (l_c+1) \phi + i \frac{1}{\sqrt{k}}
  m_c (X_L - X_R ) } (u_c)
 \right \rangle ^\text{sL} ~. \nonumber
\end{align}
On the cigar side we have $N$ bulk primary operators $\Psi^{j_a}_{m_a , \bar m_a}$ in the positions $z_a$. These have representation labels $m,\bar m,j$ where the discrete labels $m,\bar m$ are related to momentum and winding along the circular direction, and the continuous label $j$ is related to momentum in the semi-infinite direction. The boundary conditions in the cigar corresponds to a single D1-brane. There are $M$ boundary operators $\Psi^{l_c, i_c}_{m_c }$ located at the points $u_c$. These have labels $m,l,i$. Here $m$ describes the winding in the circular direction which can be half-integer (see figure \ref{fig:windingstring} below), and $l$ is again momentum in the non-compact direction. Further $i=0,1,2,3$ labels an SU(2) Chan-Paton factor \eqref{CPmatrices} which will be important for us. It is related to which branches the string ends are attached to. There is a single coupling constant $k$ in the cigar model which is the square of the cigar radius at infinity. The precise definitions of the above bulk and boundary operators will be given in eqs.
\eqref{bulkp}--\eqref{mw}.

On the sine-Liouville side the theory is described by two scalars $\phi,X$ which are non-compact and compact respectively. There are also $N$ bulk fields and $M$ boundary fields which will depend on $k$ under the mapping. The boundary conditions correspond to a D2-brane (after T-dualizing the $X$ direction), and the boundary operators will also have Chan-Paton factors which are traced over in the evaluation of the correlator. The action will depend on the coupling $1/b=\sqrt{k-2}$, and the exact form of the boundary action will be derived, see eq. \eqref{nbdly2}. Finally, there is a constant, $\mathcal N$, relating the two correlators, which only depends on $N,M$, the coupling $k$ and the total winding number.

\subsection{D1-branes in the 2d black hole}

The derivation of the duality follows the method used in \cite{HS3} which is based on path integral techniques. We thus need the boundary action for D1-branes in the 2d black hole. This is not known yet, so we first have to find it.
The starting point is the $H_3^+$ WZNW model, where $H_3^+=\mathrm{SL(}2,\mathbb C)/\mathrm{SU(}2,\mathbb R)$ is the Euclidean version of AdS$_3$.
The sigma model of the 2d black hole is the $H_3^+$ WZNW model gauged by $\mathbb R$, and
it can be embedded in the $H_3^+\times U(1)$ WZNW model \cite{Gawedzki:1991yu}.
Branes in the cigar then descend from branes in the WZNW model \cite{RS},
and it was found that there are D0-, D1- and
D2-branes in the 2d black hole.
The D0-branes descend from fuzzy spherical branes in $H_3^+$, the D1-branes from
AdS$_2$ branes, and the D2-branes from $H_2^+$ branes.
We consider D1-branes, since AdS$_2$ branes have a nice Lagrangian description \cite{FR}.
This allows us to find an action for D1-branes as we will now explain.
In particular, we will find that a Chan-Paton factor should be included in the boundary action.

Let us start with the bulk theory before going into the details of the boundary theory.
By the standard technique the coset model $H_3^+/\mathbb{R}$
can be described by the product of
the $H_3^+$ model, a U(1) free boson $X$, and a $(b,c)$-ghost system.
The action of the bulk 2d black hole is then given by%

\begin{align}
 S = S^H
 + \frac{1}{ \pi} \int d^2 w \partial X \bar \partial X
 \, \left ( + S_{b,c} \right ) ~,
 \label{cigaction}
\end{align}
where we use the action for the $H_3^+$ model in the free field realization as%
\footnote{Here the measure is $d^2 w = d x dy$ with $w = x + i y$.
Thus there is a factor 2 difference from the one in \cite{HS3}.
The world-sheet metric and its curvature are given by
$d s^2 = |\rho (w) |^2 dw ^2$ and
$\sqrt{g} {\cal R} = - 4 \partial \bar \partial \ln |\rho |$.
The regularization at the same position is done as
$\lim_{w \to z} | w - z |^2 = - \ln | \rho (z) |^2$.
In this note we set $\rho = 1$ and suppress the curvature terms.
}
\begin{align}
S^H = \frac{1}{ \pi} \int d^2 w
\left ( \partial \phi \bar \partial \phi - \beta \bar \partial \gamma
 - \bar \beta \partial \bar \gamma + \frac{Q_\phi}{4} \sqrt{g} {\cal R} \phi
 -  \pi \lambda \beta \bar \beta e^{2 b \phi}
\right) ~. \label{Haction}
\end{align}
The theory consists of a $(\beta,\gamma)$-system with conformal dimension
$(1,0)$ and a free boson $\phi$ with background charge
$Q_\phi = b = 1/\sqrt{k - 2}$.
As mentioned in the introduction we set \mbox{$\alpha ' = 1$}.
Namely, the operator products are given by
$\phi(z,\bar z) \phi(0,0) \sim X(z,\bar z) X(0,0) \sim - \ln |z|$.
Since the $(b,c)$-ghost system does not appear in the following discussion,
we neglect the part associated with the action $S_{b,c}$.

\begin{figure}
\begin{center}
\includegraphics[scale=0.7, angle=-90]{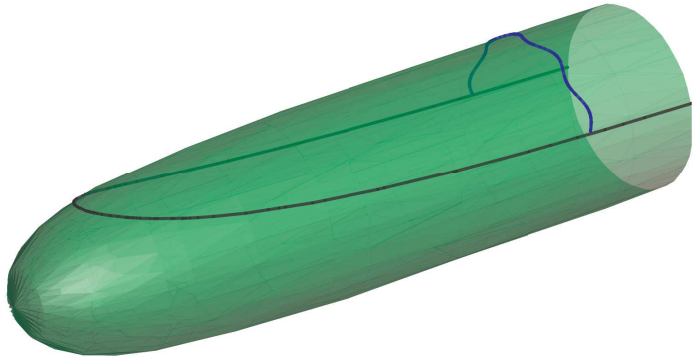}
\caption{{\em The cigar with a D1-brane (drawn in black) having $r\neq0$ and a $1/2$ winding string (drawn in blue) stretched between its two branches. $\rho$ is the coordinate along the cigar axis, and $\theta$ the angular coordinate. The tip of the cigar is located at $\rho=0$. The brane parameter $r$ describes the minimal $\rho$-value for the brane, and the parameter $\theta_0$ its angular orientation.}}\label{fig:windingstring}
\end{center}
\end{figure}

D-branes in the cigar model were investigated in \cite{RS}.
The semi-classical analysis was done by using DBI actions
as world-volume theories. Moreover, exact solutions were
obtained by making use of the fact that correlation functions on a disk are
given by the product of those in the $H_3^+$ model and the U(1)
free boson.
In this paper we consider D1-branes,
and their classical geometry can be examined as follows.
For D1-branes, it was argued that we should assign boundary
conditions corresponding to AdS$_2$ branes in the $H_3^+$
model, and Dirichlet boundary condition in the U(1) direction.
It is well known that branes in WZNW models of group manifolds are described by twisted conjugacy classes \cite{Alekseev:1998mc}.
The situation is similar in a coset, as $H_3^+=\mathrm{SL(}2,\mathbb C)/\mathrm{SU(}2,\mathbb R)$, as well as in a gauged WZNW model, at least if we gauge by an abelian group.
In appendix \ref{app:geometry} we explain the geometric meaning of the branes in the product theory that descend to D1-branes in the coset.
The brane of the U(1) part is just a point $e^{i\theta_0}$, i.e. it has to satisfy Dirichlet conditions. The $H_3^+$ part
is a left-translate of a twisted conjugacy class, translated by $e^{i\theta_0}$.
The metric of the  2d black hole may be given by
\begin{align}
ds^2 = k (d \rho^2 + \tanh ^2 \rho d \theta ^2 ) ~,
\end{align}
and the D1-branes are characterized by the equation
\begin{align}\label{eq:braneeq}
 \sinh \rho \sin (\theta - \theta_0) = \sinh r
\end{align}
with two parameters $r,\theta_0$ (see figure \ref{fig:windingstring}).
The parameter $r$
corresponds to the distance from the tip of cigar ($\rho=0$) to the D1-brane.
The other parameter $\theta_0$ represents the position of D1-branes
in the $\theta$-direction at $\rho \to \infty$.
In particular, the D1-brane reaches to the infinity $\rho \to \infty$ at
the two point $\theta_0,\theta_0 + \pi$.

One of the important steps to generalize the FZZ duality for a disk amplitude
is to obtain the boundary action for D1-branes in the cigar.
This is because we closely follow the method used for closed strings
in \cite{HS3}, where the path integral formulation is essential.
The D1-branes in the cigar model descend from AdS$_2$ branes
in the $H_3^+$ model. In equation \eqref{Haction} we have already written the action for the $H_3^+$ model, and the boundary action for AdS$_2$ branes is proposed in \cite{FR} as
\begin{align}
 S_\text{bint} = i \lambda_B  \int d u \beta e^{b \phi} ~,
 \label{hbint}
\end{align}
where the parameter $\lambda_B$ is related to $r$ as
\begin{align}
 \lambda_B = \sqrt{  \frac{ \lambda}{\sin \pi b^2}}
 \sinh r ~.
 \label{lambdaB}
\end{align}
We may treat the interaction terms perturbatively, then
the boundary conditions for free fields are Neumann
boundary condition for $\phi$, and moreover
$\gamma + \bar \gamma = 0$, $\beta + \bar \beta = 0$.
However, it turns out that this boundary action cannot be used directly for the
D1-branes in the cigar model.

We would like to propose the following modified version  as
\begin{align}
 S_\text{bint} = i \lambda_B e^{i\theta_0} \sigma_3 \int d u \beta e^{b \phi}
 \label{bint}
\end{align}
with Dirichlet boundary condition for $X$ additionally.
There are two differences from the one in \eqref{hbint}.
One is the factor $e^{i \theta_0}$, which represents the position of
the D1-brane in $\theta$-direction.
This factor is not so important, and in fact we can and will remove it by
the shift of the coordinate $\theta$.
The important one is the Chan-Paton factor $\sigma_3$, where
we use the notation for the Pauli matrices as
\begin{align}
 \sigma_0 =
  \begin{pmatrix}
   1 & 0 \\
   0 & 1
  \end{pmatrix} ~, \qquad
 \sigma_1 =
  \begin{pmatrix}
   0 & 1 \\
   1 & 0
  \end{pmatrix} ~, \qquad
 \sigma_2 =
  \begin{pmatrix}
   0 & -i \\
   i & 0
  \end{pmatrix} ~, \qquad
 \sigma_3 =
  \begin{pmatrix}
   1 & 0 \\
   0 & - 1
  \end{pmatrix} ~,
  \label{CPmatrices}
\end{align}
and $\sigma_\pm = \frac12 (\sigma_1 \pm i \sigma_2)$.
The Chan-Paton factor may be realized using boundary fermions, and
the classical consistency of \eqref{bint} is explained in appendix \ref{app:chanpaton}.

\begin{figure}
\begin{center}
\includegraphics[scale=0.3,angle=-90]{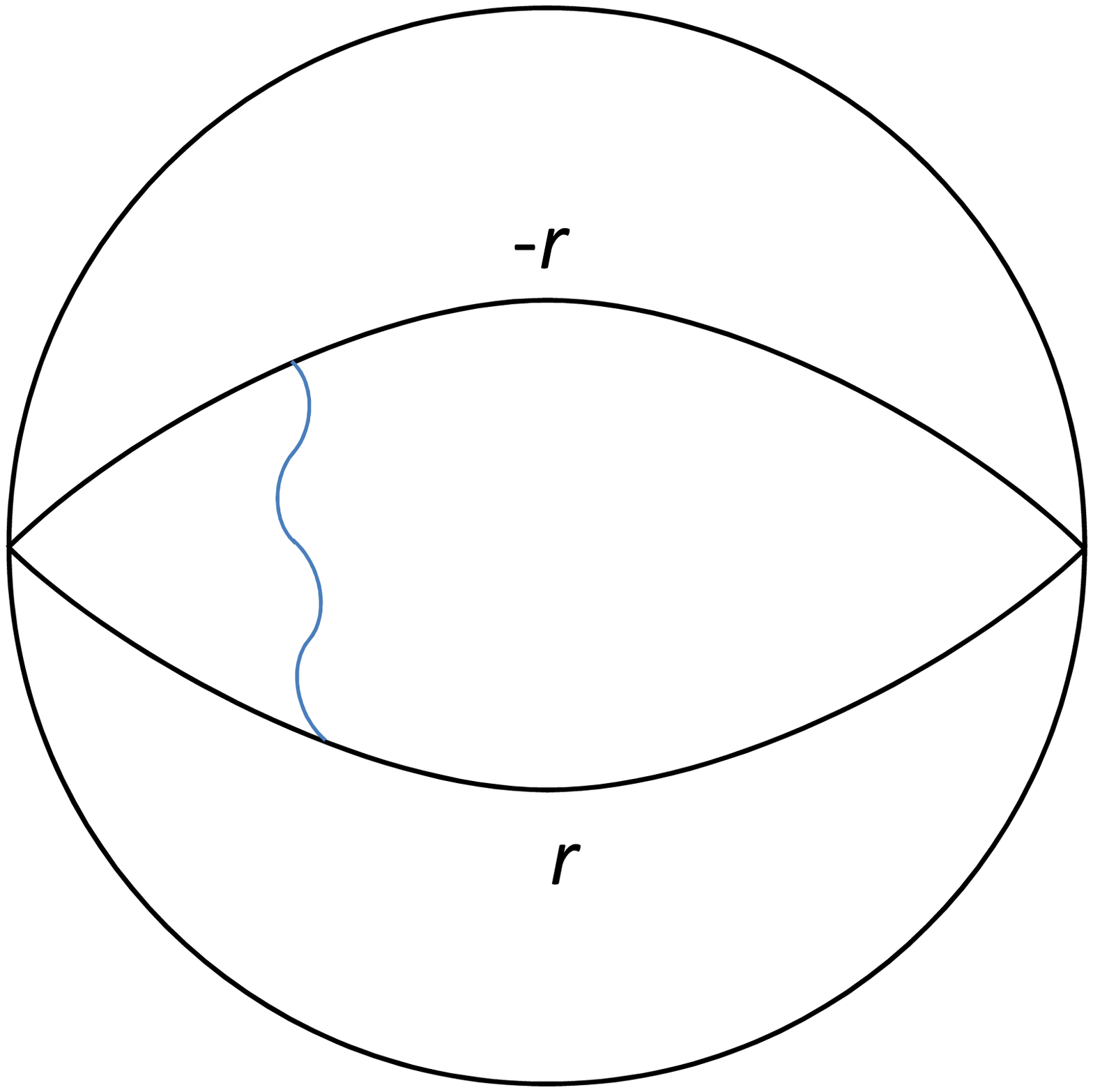}
\caption{{\em $H_3^+$ seen from above with two AdS$_2$ branes labelled $r$ and $-r$. The half integer windings strings -- sketched in blue as in figure \ref{fig:windingstring} -- now go between the two different branes.}}\label{fig:h3branes}
\end{center}
\end{figure}

We can find a reason for the Chan-Paton factor when we follow the boundary conditions from the cigar model via the gauged WZNW model to the product theory, see appendix \ref{app:gaugedsigma}. In the cigar model we can have boundary operators corresponding to half-winding open strings going between the two branches of the same D1-brane as in figure \ref{fig:windingstring}.
In fact, the spectrum of open strings between
a D1-brane was obtained in \cite{RS} with the help of modular transformation of
the annulus amplitude. It was found, for instance in eq.~(4.20) of that paper, that the
spectrum includes both integer and half-integer winding modes.
In the product theory such half-winding strings will not go between the same brane, but between two different branes located oppositely on the U(1) circle and having parameters $r$ and $-r$ for the AdS$_2$-part in $H_3^+$  as in figure \ref{fig:h3branes}.
The Chan-Paton factor appearing in the action corresponds to having these two branes, and one can explicitly see how it appears in the action when going to the product theory.
See also the discussions at the end of section 4.3 of \cite{RS}.

Let us remark that there are other ways to see the necessity of the Chan-Paton factors.
First consider the geometry of a D1-brane in the cigar which is
characterized by \eqref{eq:braneeq}. This equation is invariant
under the exchange of parameters $(r,\theta_0)$ and $(-r, \theta_0+\pi)$.
In other words, the branes with parameters $(r,\theta_0)$ and  $(-r, \theta_0+\pi)$
should not be two different branes, but identified.
In order to realize this, we take a sum of two ``fractional'' branes to construct one ``bulk'' brane in the cigar model,
just as for bulk branes in an orbifold model. This treatment is consistent with the
presence of the Chan-Paton factor $\sigma_3$. Secondly,
when we study the action, we are treating it in a perturbative way.
In the asymptotic region $\rho \to \infty$, we can neglect the interaction term
and the geometry is just a cylinder.
In this sense, the interaction terms deform the geometry in the small $\rho$ regions.
From the geometry of D1-brane \eqref{eq:braneeq}, we can see that there are
two D1-branes with opposite orientations in the $\rho \to \infty$ region.
In a flat background, it is known how to describe systems with two branes, that is, we just
need to include a $2 \times 2$ Chan-Paton factor for each open string operator.

\subsection{Correlation functions on a disk}\label{sec:corrfunconadisk}

In order to prove that the two theories are equivalent, we have to show
that all correlation functions in the two theories match.
As in \cite{HS3} we first show that correlation functions of tachyon
operators yield the same quantities.
After this we just need to check that the symmetry of the theories is the same
since descendants can be generated with the help of currents of
underlying symmetry.
First we need to develop the method to compute correlation functions on a disk
with D1-brane boundary conditions in the cigar model.
According to the general formalism, the correlators are given by
products of those of $H_3^+$ and $U(1)$ models roughly speaking,
but details should be fixed.
For instance, the boundary operators have to be associated with Chan-Paton
factors, since the boundary action includes $\sigma_3$ as we have already observed.
In the later subsections, we relate the amplitudes to
those of sine-Liouville theory.

We would like to compute correlation functions on a disk with insertions of
bulk and boundary operators.
We calculate in the product theory where the cigar model is described by the
$H_3^+\times U(1)$ model along with the ghost.
Gauge invariant bulk operators can be constructed from
the following products of operators in $H_3^+\times U(1)$
(see, for instance, \cite{HS3})
\begin{align} \label{bulkp}
 \Psi^{j}_{m , \bar m} (z)
  = \Phi^j_{m,\bar m}(z)
 V^X_{m, \bar m} (z) ~.
\end{align}
Here $\Phi^j_{m , \bar m}$ is a primary operator of $H_3^+$ model,
and it is defined as
\begin{align}
 \Phi^j_{m , \bar m} (z) = N^j_{m , \bar m} \int \frac{d ^2 \mu}{| \mu |^2}
 \mu^{m} \bar \mu^{\bar m} V_j (\mu | z) ~,
 \qquad
 V_j (\mu | z ) = | \mu |^{2j + 2} e^{\mu \gamma - \bar \mu \bar \gamma}
 e^{2 b (j+1) \phi}  \label{Hvertex}
\end{align}
with
\begin{align}
N^j_{m , \bar m} = \frac{\Gamma(-j-m)}{\Gamma (j+1+ \bar m)} ~,
\qquad m = \frac{k w + n}{2 } ~, \qquad \bar m = \frac{kw - n}{2} ~.
\end{align}
Here $n$ and $w$ take integer values and
correspond to momentum and winding number, respectively.
The $U(1)$ vertex operator
\begin{align}
  V^X_{ m, \bar m } (z)
 = e^{i \frac{2}{\sqrt{k}} ( m X_L - \bar m X_R )}
\end{align}
corresponds to a state with non-trivial winding $w$.

Along the boundary of world-sheet, we can insert boundary operators,
which are going to be constructed.
As in the bulk case, the operators should be given by certain products
of those in the $H_3^+$ and $U(1)$ models.
The boundary operators for AdS$_2$ branes in the $H_3^+$ model
were constructed in \cite{HR} as%
\footnote{
The parameter $\eta$ enters in the boundary case.
The integration over $\nu$
in \eqref{bo} has a singularity at $\nu=0$, so we have to
separate the integration domain as $\nu < 0$ and $\nu > 0$.
Alternatively, we assign the weight $\text{sgn}^\eta (\nu)$ as in \eqref{bo}.
The label $\eta$ is related to the behavior of operator under the
parity transformation $\gamma \to - \gamma$.
This can be easily seen from the fact
that the transformation $\gamma \to - \gamma$
can be absorbed by the change as $\nu \to - \nu$ and hence the
operator \eqref{bo} has a factor $(-1)^\eta$ under $\gamma \to - \gamma$.
}
\begin{align}
 \Phi^l_{m  , \eta} (u) = N^l_{m  , \eta} \int \frac{d \nu}{| \nu |}
 |\nu|^{m}  \text{sgn}^{\eta} (\nu) V_l (\nu | u) ~,
 \qquad
 V_l (\nu | u ) = | \nu |^{l + 1} e^{ (\nu \gamma - \nu \bar \gamma)/2}
 e^{b (l+1) \phi} ~.
 \label{bo}
\end{align}
Here $u$ is the coordinate of the boundary of the world-sheet
and the coefficient is
\begin{align}
 N^l_{m,\eta} = 2 i^\eta \Gamma (- l - m ) \sin \frac{\pi}{2}
 (- l - 1 - m - \eta) ~.
\end{align}
For D1-branes in the cigar model, we would like to
propose that the boundary operators are expressed as
\begin{align}
 \Psi^{l,i}_{m , \eta} (u)
  = \frac{\sigma_i}{\sqrt{2}}  \Phi^l_{m  ,\eta} V^X_{m} (u) ~,
  \label{s0v}
\end{align}
where the $U(1)$ boundary vertex operators with Dirichlet boundary condition
are
\begin{align}
  V^X_{ m } (u)
 = e^{i \frac{1}{\sqrt{k}} m (  X_L -  X_R )} ~.
\end{align}
The non-trivial point here is the Chan-Paton factor.
We have associated $\sigma_i$ to each boundary operator,
since the boundary action in \eqref{bint} includes a Chan-Paton factor $\sigma_3$.

The presence of the Chan-Paton factor affects the values of parameters
for the boundary operator.
Inserting a boundary operator changes the boundary conditions,
so it also has labels $[\Phi^l_{m,\eta}]_{L,L'}$ with
the parameters of boundary conditions $L=(r,M)$.
The parameter $r$ appears in \eqref{bint} through \eqref{lambdaB},
so the coefficient of the boundary action changes across the
boundary operator.
The parameter $M$ is related to the position of D1-brane
as $M = \frac{k}{2\pi} \theta_0$.
Since $X$ satisfies the Dirichlet boundary condition, the label $m$
represents the length of open string $\Delta \theta = 2 \pi m/\sqrt{k}$
stretched between two D1-branes at the infinity.
Thus, $m$ takes value
\begin{align}
m = M - M' + \frac{kw}{2} ~. \label{mw}
\end{align}
In order to understand the meaning of $w$, let us think of open strings on the
same brane, namely with $M=M'$. As illustrated in figure \ref{fig:windingstring}, one D1-brane reaches two
points at the infinity $\rho \to \infty$. If we associate $\sigma_i$ with $i=0,3$ in \eqref{s0v},
then the corresponding open string is stretching between the same side of the D1-brane.
Therefore, the length of open string should be $\Delta \theta \in 2 \pi \sqrt{k} \mathbb{Z}$,
which implies $w \in 2 \mathbb{Z}$ in \eqref{mw}.
In the same way, if we associate $\sigma_i$ with $i=1,2$ in \eqref{s0v},
then the corresponding open string is between the opposite sides of the D1-brane.
Therefore, the length of open string should be $\Delta \theta \in 2  \pi \sqrt{k}
(\mathbb{Z} + 1/2 )$,
which implies $w \in 2 \mathbb{Z} + 1$ in \eqref{mw}.
As argued below, we have to use
$ \eta = 0$ for $i=0,1$ and $\eta = 1$ for $i=2,3$,
so we may suppress the label $\eta$ from now on.

Let us see how the relation between the labels $\eta$ and $i$ arises.
The boundary action is given by \eqref{bint}, and the Chan-Paton factor $\sigma_3$ corresponds to having two branes. The first and the second brane have
respectively the labels $r$ and $-r$ without the Chan-Paton factor
in \eqref{bint}.
Notice that the sign change can be absorbed by $\beta \to - \beta$
and $\gamma \to - \gamma$.
First we consider the open strings
stretched between the same branes.
Then the Chan-Paton factor should be $\hat \sigma_+$
for an open string on the first brane and $\hat \sigma_-$
for an open string on the second brane. Here we have defined
$\hat \sigma_\pm = \frac{1}{2} (\sigma_0 \pm \sigma_3) $.
In other words, if we use $\sigma_0 = \hat \sigma_+ + \hat \sigma_-$,
then the open string is invariant under the exchange
of the first and the second brane. Since the exchange of branes
can be reproduced by the parity transformation
$\gamma \to - \gamma$, the open
string associated with  $\sigma_0$
should be invariant under the parity transformation.
Thus we should choose $\eta=0$ in \eqref{s0v}.
In the same way, the Chan-Paton factor $\sigma_3$
implies a minus sign under the parity transformation, and hence $\eta=1$.
Open strings between different branes can be analyzed in
the same way. The Chan-Paton factor $\sigma_1$ is invariant
under the exchange of branes by definition, and thus the
corresponding open string should be invariant under the parity
transformation. On the other hand, the Chan-Paton factor $\sigma_2$
yields a minus sign under the exchange, and we should choose
$\eta = 1$ in \eqref{s0v}.

Now that bulk and boundary operators are constructed,
we can write down the correlation functions.
We would like to compute a disk amplitude with D1-brane
boundary conditions and with the insertions of
$N$ bulk and $M$ boundary operators
\begin{align}
\left \langle \prod^N_{a = 1} \Psi^{j_a}_{m_a , \bar m_a} (z_a)
  \prod^M_{b = 1} \Psi^{l_b, i_b}_{m_b } (u_b)
 \right \rangle ~.
 \label{cf0}
\end{align}
The total winding number
\begin{align}
\sum_a ( m_a + \bar m_a ) + \sum_b m_b = \frac{kS}{2}
\end{align}
is not necessarily zero due to the cigar-shape background and
U-shape brane geometry.
As we will see below, the violation number is limited as $|S| \leq 2N+M-2$.
If it is non-zero, then the spectral flow
operation in the $H_3^+$-sector is important as emphasized in \cite{HS3}.

Let us first recall the operator inducing the spectral flow action with $S'$ for the bulk case,
which is represented by ${v}^{S'}(\zeta)$. Here we assume $S' \geq 0$, but $S' < 0$ case can
be analyzed in a similar way.
The insertion of ${v}^{S'}(\zeta)$ has two effects in our free field realization of the
$H_3^+$ model with the action \eqref{Haction}.
One is the insertion of a vertex operator $\exp(S' \phi /b)$.
The other is the restriction of the integration domain for $\beta, \bar \beta$
such that they have a zero at $z=\zeta$ of order $S'$.
See \cite{HS3} for the relation to the usual definition of spectral flow action.
For the boundary case, we would like to introduce a boundary operator
which induces the spectral flow action with $S$ units.
Here we again assume that $S \geq 0$.
We propose that it is given by $\sigma_1^S v^S(\xi)$.
The action of $v^S(\xi)$ is almost the same as the bulk case.
Namely, the boundary operator  $\exp (S \phi /2b)$ is inserted,
and the integration domain for $\beta = - \bar \beta$ is restricted
such that it has a zero at $z=\xi $ of order $S$.
The boundary operator should be associated with a Chan-Paton
factor $\sigma_i^S$. {}From the parity property, we should choose $i=0$
or $i=1$. Among them we adopt $i=1$ since the
spectral flow with odd $S$ maps open strings between the
same brane to the branes with $r$ and $-r$, and vise versa
as mentioned in \cite{RS}.

With the preparation of spectral flow operator,
 the correlation function \eqref{cf0} is now written
in terms of  $H_3^+$ and $U(1)$ models as
\begin{align}
\label{cf}
 &\left \langle \prod^N_{a = 1} \Psi^{j_a}_{m_a , \bar m_a} (z_a)
  \prod^M_{b = 1} \Psi^{l_b, i_b}_{m_b } (u_b)
 \right \rangle  \\  \nonumber
 &= \prod^N_{a = 1}  \left [ N_{m_a , \bar m_a}^{j_a}
  \int \frac{d^2 \mu_\nu}{| \mu_\nu |^2} \mu_\nu^{m_\nu}
 \bar \mu_\nu^{\bar m_\mu } \right ]
\prod^M_{b = 1}  \left [ N_{m_b , \eta_b}^{l_b}
  \int \frac{d \nu_b}{| \nu_b | } | \nu_b |^{m_b}
 \text{sgn}^{\eta_b} ( \nu_b ) \right ] \\
  & \times \text{tr} \, P \left \langle \sigma_1^S
 v^S (\xi )V^X_{ - \frac{kS}{2}} (\xi ) \prod^N_{a = 1} V_{j_a} (\mu_a | z_a)
  V^X_{m_a , \bar m_a } (z_a ) \prod^M_{b = 1} \frac{\sigma_{i_b}}{\sqrt2}
  V_{l_b} (\nu_b | u_b) V^X_{m_b } (u_b )
 \right \rangle ~, \nonumber
\end{align}
where $P$ represents the path ordering of boundary operators.
The spectral flow action $\sigma_1^S v^S (\xi) $ is inserted by
using the fact that the identity operator in the cigar model can be
represented by
\begin{align}
\mathbf{1} = \sigma_1^S v ^S (\xi) V^X_{ - \frac{kS}{2}} (\xi ) ~.
\end{align}
This is actually the very definition of the spectral flow operator.
Total winding number in the $U(1)$ model is now conserved due to
the insertion of $V^X_{ - kS/2} $.
Notice that the end result would not depend on the position
of the insertion $\xi$
since we have just inserted the identity operator.
For more details, see \cite{HS3}.

\subsection{Relation to Liouville field theory}

The task is now to express the $N+M$ point correlation
function \eqref{cf0} of the boundary cigar in
terms of a $N+M$ point correlator of sine-Liouville theory.
As in the bulk case \cite{HS3}, this can be achieved in two steps;
The first one is to rewrite the correlator in terms of a correlation function in Liouville field theory plus a free boson with additional degenerate field insertions. This is the subject of this subsection.
For this purpose we use the relation between correlators of
$H_3^+$ model and Liouville field theory \cite{RT,HS}, which was
extended to the case with boundary in \cite{HR,FR}.
The second step is to show that after applying
the self-duality of Liouville field theory,
rotations of the fields and reflection relations of operator
lead to the correlator in sine-Liouville theory.
This will be done in the next subsection.

As in \cite{HS3}, we use the path integral form of
the $N+M$ point correlation function \eqref{cf0} as
\begin{align}
&\left \langle \prod^N_{a = 1} \Psi^{j_a}_{m_a , \bar m_a} (z_a)
  \prod^M_{b = 1} \Psi^{l_b, i_b}_{m_b } (u_b)
 \right \rangle  \nonumber \\
  & \qquad= \text{tr} \, P \int {\cal D} \phi {\cal D}^2 \beta {\cal D} ^2 \gamma {\cal D} X
   e^{- S} \prod^N_{a = 1} \Psi^{j_a}_{m_a , \bar m_a} (z_a)
  \prod^M_{b = 1} \Psi^{l_b, i_b}_{m_b } (u_b) ~.
  \label{cigpath}
\end{align}
Here the action $S$ is given by \eqref{cigaction} with the boundary interaction term
\eqref{bint}.
Now we work on the upper half plane $\text{Im}\, z \geq 0$ and the boundary
is at the line $\text{Im}\, z = 0$. According to the standard doubling trick,
the anti-holomorphic part of fields are mapped to the region of
$\text{Im}\, z < 0$ such that the boundary conditions are satisfied along the
boundary  $\text{Im}\, z = 0$.

Following \cite{HS,FR}, we integrate over $\beta,\gamma$ to
reduce the theory to the one with two remaining fields $\phi,X$.
Notice that the field $\gamma$ appears only linearly in the exponent of
the path integral \eqref{cigpath}. This is because the action includes $\gamma$
only in the kinetic term of \eqref{cigaction} and the vertex operators are expressed
in terms of \eqref{Hvertex} and \eqref{bo}.
Therefore, by integrating over $\gamma$, we would have a delta functional for
$\bar \partial \beta$, which implies that the field $\beta$ can be replaced by a function
${\cal B} (z)$. Integrating over the world-sheet coordinate, the function is obtained
as
\begin{align}
\label{calB}
 {\cal B}(w) = \sum_{a=1}^N \frac{\mu_a}{w - z_a}
 + \sum_{a=1}^N \frac{\bar \mu_a}{w - \bar z_a}
 + \sum_{b=1}^M \frac{\nu_b}{w - u_b} ~.
\end{align}
In the same way, $\bar \beta (\bar z) $ is replaced by $- \bar {\cal B} (\bar z)$.
This form might be understood from the operator product between $\beta$ in the interaction
terms and $\gamma, \bar \gamma$ in the vertex operators. See also \cite{FR}.

One essential ingredient of the $H_3^+$-Liouville relation is the change of variables
corresponding to Sklyanin's separation of variables as in \cite{RT,HS}.
With the boundary, the formula for the change of variables is given by
(see \cite{HR,FR} and also appendix \ref{Sklyanin})
\begin{align}
{\cal B}(w)  = u \frac{(w - \xi)^S \prod_{a' = 1}^{N '} (w - y_{a'}) (w - \bar y_{a'})
   \prod_{b' = 1}^{M '}  (w - t_{b'})  }
    {\prod_{a = 1}^{N } (w - z_{a}) (w - \bar z_{a})
   \prod_{b = 1}^{M }  (w - u_{b}) } ~.
   \label{calB2}
\end{align}
Here $t_{b'}$ and $y_{a'}$ denote zeros respectively on the boundary and in the bulk.
The equality gives a map of parameters from $(\mu_a , \bar \mu_a , \nu_b)$
to $(u, y_{a'}, \bar y_{a'} , t_{b'})$, where the numbers of $y_{a '}$ and $t_{b'}$
satisfy the equation $2 N ' + M' + S + 2 = 2 N + M$.
This follows from ${\cal B} (z)$ being a meromorphic one-form on the full plane hence having two more poles than zeros.
Notice that the numbers $N', M'$ would vary when we change the
values of $(\mu_a , \bar \mu_a , \nu_b)$.
Moreover, since the numbers of insertions should be non-negative, we have
a restriction for the total winding number $S \leq 2N + M - 2$.
The presence of $v^S(\xi)$ forces $\beta$ to have a zero of order $S$, and this is possible only if
\begin{align}
 \ell_n (\xi ) = \sum_{a=1}^N \frac{\mu_a}{( \xi - z_a)^n }
 + \sum_{a=1}^N \frac{\bar \mu_a}{ ( \xi  - \bar z_a )^n }
 + \sum_{b=1}^M \frac{\nu_b}{ ( \xi - u_b )^n} = 0
 \label{constmu}
\end{align}
for $n = 0,1,\cdots , S$.
Due to the  $S+1$ constraints, the number of parameters matches.
Since the correlation function is given by the integration over
 $(\mu_a , \bar \mu_a , \nu_b)$ in \eqref{cf}, we need the formula for the
Jacobian due to the change of variables.
It is given by
\begin{align}
 \int \prod_{a=1}^N\frac{d^2 \mu_a}{|\mu_a|^2}
 \prod_{b=1}^b\frac{d \nu_b}{|\nu_b|} \prod_{n=0}^S \delta (\ell_n ( \xi ))
  = \sum_{N' , M'} \frac{1}{N'! M'!} \int  \frac{d u }{|u|^{2+S}} \prod_{a ' = 1}^{N '} d^2 y_{a '}
    \prod_{b ' =1}^{M '} d t_{b '}  | \Xi |
    \label{jacobian}
\end{align}
with
\begin{align}
 \Xi &=  \prod_{i < j} |z_{ij} |^2 \prod_{i , j} ( z_i - \bar z_j )
  \prod_{i , a} | z_i - u_a |^2 \prod_{a < b} u_{ab}
    \prod_{i '< j'} |y_{i'j'} |^2 \prod_{i' , j'} ( y_{i'} - \bar y_{j'} )
  \prod_{i' , a'} | y_{i'} - t_{a'} |^2 \prod_{a' < b'} t_{a'b'}
  \nonumber \\
  &\times \prod_{i , i'} |z_i - y_{i'}|^{-2}
  \prod_{i,i'}|z_i - \bar y_{i'}|^{-2}
   \prod_{i, a'} |z_i - t_{a'}|^{-2}
   \prod_{a , i'} | u_a - y_{i'} |^{-2}
   \prod_{a, a'} (u_a - t_{a'})^{-1} ~.
   \label{theta}
\end{align}
Here we should sum over $N',M'$ with the condition $2 N ' + M ' + S + 2 = 2 N + M$
since the numbers $N',M'$ depend on the explicit values of $(\mu_a , \bar \mu_a , \nu_b)$.
For details, see appendix \ref{Sklyanin}.

The action of $H_3^+$ model is given by \eqref{Haction}, and now $\gamma (w)$
are integrated over and $\beta (w)$ is replaced by a function ${\cal B}(w)$.
Now the theory is like the Liouville field theory, but the interaction term includes
coordinate dependent coefficients as $|{\cal B} (w) |^2 \exp (2 b \phi (w))$.
As in \cite{HS,HS3}, we change the field $\phi$ as
\begin{align}
\varphi(w,\bar w)=\phi(w,\bar w)+\frac{1}{2b}\ln |{\cal{B}}(w)|^2  ~,
\label{changephi}
\end{align}
then the interaction term becomes $\exp (2 b \varphi (w))$ as desired.
The boundary interaction term \eqref{bint} is now
\begin{align} \label{bintm}
 S_\text{bint} = i \lambda_B  \sigma_3
\int d t \, \text{sgn} \, {\cal B } (t)e^{b \varphi }  ~,
\end{align}
which has a  sign function $\text{sgn} \, {\cal B } (t)$ as noticed
in \cite{HR,FR}. This implies that the parameter of boundary interaction
changes when it crosses the boundary positions $u_b$ and $t_{b'}$.
As noticed in \cite{HS3}, it is also necessary to shift
$X (w, \bar w ) = X_L (w) + X_R (\bar w)$ as
\begin{align}
  \chi_L(w)=X_L(w)-i\frac{\sqrt{k}}{2}\ln {\cal B} (w)~, \qquad
  \chi_R (\bar w) = X_R (\bar w) + i \frac{\sqrt{k}}{2}\ln \bar {\cal B} (\bar w) ~,
  \label{changex}
\end{align}
where $\chi_R(\bar w)$ is the complex conjugate of $\chi_L(w)$.
This means that $X=X_L+X_R$ is changed with $-i\sqrt{k}/2\ln \cal{B}/ \cal{\bar B}$ whereas the dual field $\tilde X=X_L-X_R$ is changed with $-i\sqrt{k}/2\ln |{\cal{B}}|^2 $.
The action is now
\begin{align}
 S = \frac{1}{ \pi} \int d^2 w \left(
 \partial \varphi \bar \partial \varphi + \partial \chi \bar \partial \chi
 + \frac{\sqrt{g}}{4}
 {\cal R} (Q_\varphi \varphi + Q_{\tilde \chi} \tilde \chi )
 +  \pi \lambda  e^{2 b \varphi}  \right)
 \label{LLaction}
\end{align}
with boundary interaction term \eqref{bintm}.
The dual field is denoted by $\tilde \chi = \chi_L - \chi_R$ and
background charges are found to be shifted as $Q_\varphi = b + b^{-1}$ and
$Q_{\tilde \chi} = - i \sqrt{k}$ as shown in \cite{HS,HS3}.

The change of fields \eqref{changephi} and \eqref{changex} also affects the kinetic terms.
We write the action as $- \int d^2 w \phi \partial \bar \partial \phi$
and insert the expression in \eqref{changephi}. Then, from the term
$\partial \bar \partial \ln | {\cal B } (w) |^2$, we obtain delta functions
localized at $z_a,u_b,y_{a'},t_{b'}$. Integrating over the world-sheet coordinate, we
find shifts of momenta in the existing vertex operators at $z_a,u_b$ and insertions of
new operators at $y_{a'},t_{b'}$, since the action is in the exponent of the path integral
\eqref{cigpath}. The similar things happen for $X$ due to the shift \eqref{changex}.
Closely following the analysis in \cite{HS,HS3}, the correlation function
\eqref{cf0} is now written as
\begin{align}  \label{corrmid}
 &\left \langle \prod^N_{a = 1} \Psi^{j_a}_{m_a , \bar m_a} (z_a)
  \prod^M_{b = 1} \Psi^{l_b, i_b}_{m_b } (u_b)
 \right \rangle
   =  \sum_{M' , N'} \frac{1}{N' ! M' !}\prod^{N '}_{a ' = 1} \int d^2 y_{a '}
     \prod^{M '}_{b ' = 1} \int  d t_{b '} \\
       & \qquad  \times
  \prod^N_{a = 1}  N_{m_a , \bar m_a}^{j_a}
\prod^M_{b = 1}   N_{m_b , \eta_b}^{l_b}  \text{sgn}^{\eta_b} ({\cal B} (u_b))
 \, \text{tr} \, P \left \langle \sigma_1^S \prod^N_{a = 1} V_{\alpha_a} ( z_a)
  V^\chi_{m_a - \frac{k}{2}, \bar m_a - \frac{k}{2}} (z_a )
   \right. \nonumber \\ & \qquad  \times \left.
 \prod^M_{b = 1}
  B^{i_b}_{\beta_b} ( u_b) V^\chi_{m_b - \frac{k}{2} } (u_b )
  \prod^{N ' }_{a ' = 1} V_{- \frac{1}{2b}} ( y_{a'})
  V^\chi_{ \frac{k}{2}, \frac{k}{2}} (y_{a'} )
 \prod^{M '}_{b ' = 1}
  B^0_{- \frac{1}{2b}} ( t_{b'}) V^\chi_{ \frac{k}{2} } (t_{b'} )
  \right \rangle ~. \nonumber
\end{align}
The right hand side is computed using the action \eqref{LLaction}
with the boundary interaction term \eqref{bintm}.
The bulk and boundary operators in the boundary Liouville field theory are defined as
\begin{align}
V_\alpha (z) = e^{ 2 \alpha \varphi (z) }  ~, \qquad
B^i_\beta (t) = \frac{\sigma_i}{\sqrt2} e^{ \beta \varphi (t)} ~,
\end{align}
respectively. The shifts of parameters for bulk and boundary operators at $z_a,u_b$ are
\begin{align}
\alpha_\nu = b(j_\nu + 1) + \frac{1}{2b} ~, \qquad
\beta_\nu = b(l_\nu + 1) + \frac{1}{2b} ~,
\end{align}
and new operators are inserted at $y_{a'},t_{b'}$.
In the above expression ${\cal B} (u_b)$ is actually not well-defined since
${\cal B} (z)$ has a pole at $z=u_b$. We just represent
$ \text{Res}_{z \to u_b} {\cal B}(z) (= \nu_b)$ by ${\cal B} (u_b)$ for simplicity.
{}From the terms like $\ln |{\cal B}(w)| \partial \bar \partial \ln |{\cal B}(w)|  $,
we would obtain a pre-factor in the correlation function, but as found in \cite{HS3}
we can see that it cancels the Jacobian factor \eqref{theta}.
Moreover, the insertion $\exp (S \phi /2b)$ coming form $v^S (\xi)$
disappears due to the shift of momenta at $\xi$.

The method in \cite{HS3} can be applied almost straightforwardly,
one exception is, however, the insertion of the new boundary operator
\begin{align}
B^0_{- \frac{1}{2b}} (t_{b'}) = \frac{\sigma_0}{\sqrt2} e^{ - \frac{1}{2b} \varphi (t_{b'}) }
\label{b0d}
\end{align}
with a non-trival factor $\sigma_0/\sqrt2$.
If we follow the analysis in \cite{HS,FR},
then we obtain the insertion of $e^{ - \varphi (t)/2b}$
along the boundary. However, it is required to associate a Chan-Paton factor to the insertion at the boundary.
The factor should be proportional to the identity operator $\sigma_0$,
but the overall normalization may differ from one.
Let us first examine the operator corresponding to the identity state.
The identity state $| 1 \rangle$ has the norm
$\langle 1|1 \rangle =1$. On the other
hand, the identity operator ${\cal O}_1$ is defined as
${\cal O}_1 \cdot {\cal O}= {\cal O} $ for all
operators $\cal O$. Due to the state-operator correspondence, we
may define $|1 \rangle := C {\cal O}_1 |0 \rangle $
up to an overall factor $ C$.
For our case ${\cal O}_1 = \sigma_0$, and
  $\langle 1|1 \rangle = C^2 \, \text{tr} \, \sigma_0^2
\langle 0|0 \rangle
= 2 C^2$.
Therefore, we should define as $|1 \rangle := \sigma_0 /\sqrt{2} |0 \rangle$,
and the operator corresponding to the identity
state is $ \sigma_0/ \sqrt{2}$.
This reasoning also explains the factor $\sigma_0/\sqrt2$ in \eqref{b0d}.

The aim of this subsection was to rewrite the correlation function
of the cigar model in terms of Liouville field theory and free boson
with the action \eqref{LLaction}, \eqref{bintm} as we have already
done in \eqref{corrmid}.
However, in order to proceed furthermore, it is convenient to remove the
sign factors  $\text{sgn} \, {\cal B} (t)$ in the boundary action
\eqref{bintm} and $\text{sgn}^{\eta_b} ({\cal B} (u_b))$ in front of the
correlator \eqref{corrmid}.
This is possible by making use of the anti-commutativity of the Pauli
matrices.
Since the boundary action \eqref{bintm}
includes $\sigma_3$, the boundary operator
$B^i_\beta (t)$ commutes with it for $i=0,3$ and anti-commutes with it
for $i=1,2$. As mentioned below equation \eqref{bintm}, the role of $\text{sgn} \, {\cal B} (t)$
is to multiply $(-1)$ when it crosses the positions of boundary operators.
Therefore, we can remove the sign function by replacing $i=0,3$ and $i=1,2$.
Moreover, the other sign functions  $\text{sgn}^{\eta_b} ({\cal B} (u_b))$
implies that we receive  extra minus sign when boundary operators
with $\eta_b = 0$ and $\eta_b = 1$ are exchanged.
Recalling that $i=0,1$ for $\eta = 0$ and $i=2,3$ for $\eta = 1$,
the effect of $\text{sgn}^{\eta_b} ({\cal B} (u_b))$ can be reproduced
by replacing $i=0$ and $i=1$ regardless of the replacement of
$i=2$ and $i=3$.
Combining the both, the rule may be summarized such that
$\sigma_i$ in front of the boundary operator is replaced by
$\sigma_{1} \sigma_i$.
Effectively we may insert
$1 = \sigma_1^{M + M' + S}$ with even $M + M' + S$ to the correlation
function.
Now the expression becomes a bit simpler%
\footnote{The equality is up to a trivial factor.}
\begin{align}
\label{ctol}
 &\left \langle \prod^N_{a = 1} \Psi^{j_a}_{m_a , \bar m_a} (z_a)
  \prod^M_{b = 1} \Psi^{l_b, i_b}_{m_b } (u_b)
 \right \rangle
   =   \sum_{M' , N'} \frac{1}{N' ! M' !}
    \prod^{N '}_{a ' = 1} \int \frac{d^2 y_{a '}}{N' !}
     \prod^{M '}_{b ' = 1} \int  \frac{d t_{b '}}{M' !} \\
       & \qquad  \times
  \prod^N_{a = 1}  N_{m_a , \bar m_a}^{j_a}
\prod^M_{b = 1}   N_{m_b , \eta_b}^{l_b}
 \, \text{tr} \, P \left \langle  \prod^N_{a = 1} V_{\alpha_a} ( z_a)
  V^\chi_{m_a - \frac{k}{2}, \bar m_a - \frac{k}{2}} (z_a )
   \right. \nonumber \\ & \qquad  \times \left.
 \prod^M_{b = 1} \sigma_{1}
  B^{i_b}_{\beta_b} ( u_b) V^\chi_{m_b - \frac{k}{2} } (u_b )
  \prod^{N ' }_{a ' = 1} V_{- \frac{1}{2b}} ( y_{a'})
  V^\chi_{ \frac{k}{2}, \frac{k}{2}} (y_{a'} )
 \prod^{M '}_{b ' = 1}
  B^1_{- \frac{1}{2b}} ( t_{b'}) V^\chi_{ \frac{k}{2} } (t_{b'} )
  \right \rangle  \nonumber
\end{align}
with boundary interaction term
\begin{align}
 S_\text{bint} = i \lambda_B  \sigma_3
\int d t  \, e^{b \varphi }  ~.
\end{align}
The function ${\cal B}(w)$ is proportional to $u$ as
defined in \eqref{calB}, but the dependence of $u$ disappears in the
last expression. For instance, the dependence of $\text{sgn} (u)$
can be absorbed by the rotation of Chan-Paton factors as
$\sigma_2,\sigma_3 \to - \sigma_2 , - \sigma_3$.

\subsection{Duality with boundary sine-Liouville theory}

In the previous subsection, we obtained a relation between the cigar model
and a Liouville-like theory, but it is not what we wanted to have.
In order to relate the cigar model to the sine-Liouville theory,
we take three steps as in \cite{HS3}.
First step is to perform the self-duality
of the Liouville theory exchanging
$b \leftrightarrow b^{-1}$. With this duality the relation
becomes a strong/weak duality for $k$.
Next step is to realize that the extra insertions at $y_{a'}, t_{b'}$
can be seen as an expansion of an interaction term in the action. Thus we obtain a relation between
$N + M$ point correlation functions. The final step is to perform a rotation of fields and utilize the reflection
relation such as to arrive at the boundary sine-Liouville theory.

First step is the Liouville self-duality.
It is known that the Liouville field theory is self-dual under the exchange of
$b$ by $b^{-1}$ followed by the replacement of
$\lambda$ by $\tilde \lambda$ as
\cite{Zamolodchikov:1995aa}
\begin{align}
 {\cal L} = \tilde \lambda e^{2 b^{-1} \varphi} ~, \qquad
  \pi \tilde \lambda \gamma (1/b^2) = (\pi \lambda \gamma (b^2))^{1/b^2} ~,
\label{dpara}
\end{align}
where $\gamma(x) = \Gamma(x)/\Gamma(1-x)$.
The self-duality is extended to the case with boundary, and
the dual interaction term is given by \cite{FZZ}
\begin{align}
 {\cal L}_B = \tilde \lambda_B  e^{ b^{-1} \varphi} ~, \qquad
  \lambda_B = \sqrt{\frac{ \lambda}{\sin \pi b^{2}}}
  \cosh b s ~, \qquad
 \tilde \lambda_B = \sqrt{\frac{\tilde \lambda}{\sin \pi b^{-2}}}
  \cosh b^{-1} s
  \label{formulafzz}
\end{align}
for the case without Chan-Paton factor.
In our case, it is convenient to rewrite the Pauli-matrix as
$\sigma_3 = \sigma_+ \sigma_- - \sigma_- \sigma_+$,
 such that the
interaction term can be treated as two single branes.
Applying the formula \eqref{formulafzz} for each term with
$ \sigma_+ \sigma_-$ and $\sigma_- \sigma_+$,
the dual interaction term for the boundary is obtained as
\begin{align}
 {\cal L}_B = f (\sigma_i) e^{b^{-1} \varphi }  ~,
\end{align}
with
\begin{align}
 f (\sigma_i) &= \sqrt{\frac{\tilde \lambda}{\sin \pi b^{-2}}}
   \left( \cosh \left( b^{-2} \left(r + \frac{\pi i}{2} \right) \right) \sigma_+ \sigma_-
 +  \cosh \left( b^{-2} \left(r - \frac{\pi i}{2} \right) \right)  \sigma_- \sigma_+ \right) \\
  &= \sqrt{\frac{\tilde \lambda}{\sin \pi b^{-2}}}
   \left( \cosh  \frac{r}{b^2} \cos \frac{\pi}{2 b^2} \sigma_0
    + i \sinh \frac{r}{b^2} \sin \frac{\pi}{2 b^2} \sigma_3 \right) ~.
    \nonumber
\end{align}
The second equality comes from $\sigma_\pm \sigma_\mp = \frac12 (\sigma_0 \pm \sigma_3)$.
The same result may be obtained with the method in appendix \ref{ra}.
With the help of the self-duality, we would obtain a strong/week duality.

Next, we focus on the extra insertions of bulk and boundary operators
\begin{align} \label{bulkint}
  V ( y_{a'}) =V_{- \frac{1}{2b}} V^\chi_{ \frac{k}{2}, \frac{k}{2}}
    = e^{- b^{-1} \varphi + i \sqrt{k} \tilde \chi } ~, \qquad
 V_B ( t_{b'})=
  B^1_{- \frac{1}{2b}}  V^\chi_{ \frac{k}{2} }
   = \frac{\sigma_1}{\sqrt{2}}
 e^{- \frac{1}{2 }b^{-1}\varphi + i \frac{\sqrt{k}}{2} \tilde \chi  } ~.
\end{align}
The $H_3^+$-Liouville relation maps the parameters $\mu_a , \nu_b$
to the positions of extra insertions $y_{a'}, t_{b'}$.
However, in the coset model, we should use the $m$-basis expressions as in \eqref{Hvertex}
and \eqref{bo} by performing Fourier transforms.
These lead to the integration over  $\mu_a , \nu_b$
and hence $y_{a'} , t_{b'}$ after the map as in \eqref{ctol}.
Since the positions of the extra insertions are integrated over the whole world-sheet $y_{a'}$
and the whole boundary $t_{b'}$,
we may deal with them as a part of the interaction terms in the action.
Notice that only the terms in \eqref{ctol} contribute after expanding
the interaction terms due to the momentum conservation along the $\chi$ direction.
The momentum conservation still allows different numbers of bulk and boundary
insertions $N',M'$ if they satisfy the condition $2 N' +  M ' = 2N + M - S - 2 $.
After the above two steps, the interaction terms for the bulk are now the sum of
\begin{align}
 {\cal L}_1 = \tilde \lambda e^{2b^{-1} \varphi} ~, \qquad
 {\cal L}_2 = - e^{- b^{-1} \varphi + i \sqrt{k} \tilde \chi } ~,
\end{align}
and for the boundary
\begin{align}
 {\cal L}_{B,1} = f (\sigma_i) e^{b^{-1} \varphi }  ~, \qquad
 {\cal L}_{B,2} = - \frac{\sigma_1}{\sqrt{2}}
 e^{- \frac{1}{2 }b^{-1}\varphi + i \frac{\sqrt{k}}{2} \tilde \chi  } ~.
\end{align}
Since the extra insertions are treated as the interaction terms, we
now have a relation between $N + M$ point correlation functions
of the cigar model and the theory with the above interaction terms.
However, the new theory is not yet the sine-Liouville field theory.

Fortunately, the new theory can be mapped to sine-Liouville theory
as in \cite{HS3}.
Since the background charges for $\varphi$ and $\chi$ are
different from those for sine-Liouville theory, we redefine
as
\begin{align}
 \phi = (k - 1) \varphi - i \sqrt{k } b^{-1} \tilde \chi ~,
 \qquad
 \tilde X = - i \sqrt{k } b^{-1} \varphi - (k -1) \tilde \chi ~.
\end{align}
The background charges then become the desired ones as
$Q_\phi = b =1/\sqrt{k-2}$ and $Q_{\tilde X} = 0$.
After this rotation the interaction terms are
\begin{align}
 {\cal L}_1 = \tilde \lambda e^{2 b^{-1}(k - 1) \phi
 - 2 i \sqrt{k} b^{-2} \tilde X} ~, \qquad
 {\cal L}_2 = - e^{ b^{-1} \phi - i \sqrt{k} \tilde X } ~,
\end{align}
and for the boundary
\begin{align}
 {\cal L}_{B,1} = f (\sigma_i) e^{ b^{-1}(k - 1) \phi
 -  i \sqrt{k} b^{-2} \tilde X} ~, \qquad
 {\cal L}_{B,2} = - \frac{\sigma_1}{\sqrt{2}}
 e^{ \frac{1}{2 } b^{-1}\phi - i \frac{\sqrt{k}}{2} \tilde X  } ~.
\end{align}
Here we should note that the rotation is consistent with the boundary
conditions since $\phi$ and $\tilde X$ satisfy Neumann boundary condition.

We also need to apply the reflection relations of Liouville theory to the
interaction terms in order to arrive at the sine-Liouville theory \cite{HS3}.
Since Liouville theory has interaction term of the exponential type,
in-coming and out-going modes are related by reflection relations.
One trick here is to treat  ${\cal L}_2$ as the
Liouville term instead of ${\cal L}_1$. More precisely speaking,
we introduce a new field $\hat \phi = - i/\sqrt{2} ( b^{-1} \phi - i \sqrt{k} \tilde X)$,
then ${\cal L}_2 = - e^{i \sqrt{2} \hat \phi}$  becomes
a Liouville interaction term with $\hat b= i/\sqrt{2}$.
We apply the reflection relation with \eqref{bulkrr} for the $\hat b= i/\sqrt{2}$ theory
to the bulk operator in ${\cal L}_1$ keeping intact the part orthogonal to the $\hat \phi$ direction.
Rotating back to the original fields, we have
\begin{align}
 {\cal L}_1 = - \tilde \lambda  \pi^{-1 - 2 b^{-2}}  \gamma (1 + b^{ - 2} )
 e^{b^{-1} \phi + i \sqrt{k} \tilde X } ~, \qquad
  {\cal L}_2 = - e^{b^{-1} \phi - i \sqrt{k} \tilde X }
  \label{nbulk}
\end{align}
for the bulk interaction terms.
They are the interaction terms of
bulk sine-Liouville theory as desired.
More detailed explanations are given in \cite{HS3}.
For the boundary case we need the reflection amplitude
with boundary interaction term
${\cal L}_{B,2} = - 2^{-1/2} \sigma_1 e^{i \hat \phi/\sqrt{2}}$,
which is studied in appendix \ref{ra}.
With the result from there, we have
\begin{align}
 {\cal L}_{B,1} =  \frac{\tilde c}{\sqrt2}  ( e^{\frac{r}{b^2}} \sigma_+
 + e^{- \frac{r}{b^2}} \sigma_-  )
 e^{\frac{1}{2 b}\phi + i \frac{\sqrt{k}}{2} \tilde X  }  ~, \qquad
 {\cal L}_{B,2} =  - \frac{1}{\sqrt2} (\sigma_+ + \sigma_-)
 e^{\frac{1}{2 b}\phi - i \frac{\sqrt{k}}{2} \tilde X  } \label{nbdly}
\end{align}
with
\begin{align}
 \tilde c  = \sqrt{ \tilde \lambda \pi^{-1-2b^{-2}}
 \gamma (1+b^{-2}) }
 ~. \label{tildec}
\end{align}
In the end of the computation we have performed a SU(2) rotation of
Chan-Paton factors as
\begin{align}
(\sigma_1,\sigma_2,\sigma_3) \to
(\sigma_1, - \sigma_3,\sigma_2)
\end{align}
just to make the expression simpler.
We would like to claim that they are the interaction terms for D2-branes
in the sine-Liouville theory.

If we want to obtain a symmetric expression, then we would shift the zero
mode of $\tilde X$ appropriately.
For the bulk interaction we have
\begin{align}
 {\cal L}_1 = \tilde c
 e^{b^{-1} \phi + i \sqrt{k} \tilde X } ~, \qquad
  {\cal L}_2 = \tilde c e^{b^{-1} \phi - i \sqrt{k} \tilde X },
  \label{nbulk2}
\end{align}
and for the boundary interaction
\begin{align}\label{nbdly2}
 &{\cal L}_{B,1} =  i \sqrt{\frac{\tilde c}{2} }  \left( e^{\frac{r}{2b^2}} \sigma_+
 + e^{- \frac{r}{2b^2}} \sigma_-  \right)
 e^{\frac{1}{2 b}\phi + i \frac{\sqrt{k}}{2} \tilde X  }  ~, \\
 &{\cal L}_{B,2} =  i \sqrt{\frac{\tilde c}{2}}
 \left(e^{- \frac{r}{2b^2}} \sigma_+ + e^{\frac{r}{2b^2}} \sigma_- \right)
 e^{\frac{1}{2 b}\phi - i \frac{\sqrt{k}}{2} \tilde X  }
~. \nonumber
\end{align}
This form of the boundary action is quite analogous to the one  for B-branes
in ${\cal N}=2$ super Liouville field theory in (5.8) with (5.27) of \cite{Hosomichi}.
{}From this fact, we are confident in the correctness of the boundary
action for D2-branes in sine-Liouville theory.
In the next section we study the fermionic version of FZZ duality, and
indeed we obtain the boundary action for a B-brane in ${\cal N}=2$
super Liouville field theory from the one for a D1-brane in the
fermionic cigar model.

We should also apply the same rotation of fields and the reflection
relations to the vertex operators in order to establish the relation between
correlation functions.
For the bulk operator it was done in \cite{HS3} as
\begin{align}
& N^j_{m , \bar m} V_{\alpha} ( z )
  V^X_{m - \frac{k}{2}, \bar m  - \frac{k}{2}} (z)
 = -  \pi^{-1-2j-m-\bar m} e^{ 2 b (j+1) \phi + i \frac{2}{\sqrt{k}}
 ( m X_L -  \bar m  X_R )} ~.
\end{align}
For the boundary operator we have in the same way
\begin{align}
 N^l_{m } \sigma_1  B^i_{\beta} ( u )
  V^X_{m - \frac{k}{2} } (u)
 =  - (-1)^\eta \pi^{ - l - m }  \sigma_i
 e^{  b (l+1) \phi + i \frac{1}{\sqrt{k}}
  m (X_L - X_R ) } ~,
\end{align}
where the pre-factor is  computed
by using the result in appendix \ref{ra}.
Here $\eta = 0$ for $i=0,1$ and $\eta = 1$ for $i=2,3$ as mentioned in subsection \ref{sec:corrfunconadisk}.
Combining all the results obtained so far, the equation \eqref{ctol}
leads to the final result mentioned already in equation \eqref{eq:finalresultbosonic}
\begin{align}\nonumber
 &\left \langle \prod^N_{a = 1} \Psi^{j_a}_{m_a , \bar m_a} (z_a)
  \prod^M_{c = 1} \Psi^{l_c, i_c}_{m_c } (u_c)
 \right \rangle  ^\text{cig}
  \\ & = {\cal N}
  \left \langle \prod^N_{a = 1} e^{ 2 b (j_a+1) \phi + i \frac{2}{\sqrt{k}}
 ( m_a X_L -  \bar m_a  X_R )} (z_a)
  \prod^M_{c = 1} \frac{\sigma_{i_c}}{\sqrt{2}}
e^{  b (l_c+1) \phi + i \frac{1}{\sqrt{k}}
  m_c (X_L - X_R ) } (u_c)
 \right \rangle ^\text{sL} ~, \nonumber
\end{align}
where the left hand side and the right hand side are computed
in the cigar model and sine-Liouville theory, respectively.

Up to now we have shown that the correlation functions for the
cigar model and sine-Liouville theory among tachyon vertex operators
are the same. For the equivalence of theory we have to establish
the relation among descendants as well. As mentioned in subsection \ref{sec:corrfunconadisk}, since the
descendants can be constructed by the action of symmetry
current generators, we just need to show the both boundary theories preserve
the same symmetry. In the cigar model, we are considering branes
preserving one pair of parafermionic currents. As shown in
\cite{Bakas:1991fs}, we can construct the generators of so-called $\hat W_\infty (k)$ algebra from the parafermionic currents.
In fact, it was already shown in \cite{FL} that the boundary actions
\eqref{nbdly2}
in sine-Liouville theory preserve the $\hat W_\infty (k)$ symmetry.

\section{Fermionic FZZ duality}

In this section we study the fermionic version of the FZZ duality
which relates the fermionic 2d black hole and ${\cal N}=2$ super
Liouville field theory. In \cite{HK} the authors show that
they are related by a mirror symmetry \cite{HV}. Here we would like to give
another proof by utilizing the method developed in \cite{HS3}.
The motivation to consider the fermionic FZZ duality is twofold.
Firstly the ${\cal N}=2$ SL(2)/U(1) coset model
or dual ${\cal N}=2$ super Liouville theory appears frequently
in the context of superstring theory. For instance, string
compactification on a singular Calabi-Yau 3-fold can be described
with these models (see, e.g., \cite{Giveon:1999px}). Second one is to check that the boundary action
\eqref{nbdly2} for D2-brane in sine-Liouville theory is the correct one.
This is possible since the counterpart in the fermionic version
is B-brane in ${\cal N}=2$ super Liouville field theory and
its boundary action has been obtained in \cite{Hosomichi}.
D-branes in ${\cal N}=2$ SL(2)/U(1) coset and
${\cal N}=2$ super Liouville theory have been also studied in, e.g.,
\cite{Eguchi:2003ik,Ahn:2003tt,Israel:2004jt,Fotopoulos:2004ut}.

\subsection{Fermionic 2d black hole}

The fermionic 2d black hole is the ${\cal N}=2$ supersymmetric model based on the coset SL(2)/U(1)
which is given by the Kazama-Suzuki construction \cite{Kazama:1988qp}.
As in the bosonic case, we start from the fermionic 2d black hole
and show that it is equivalent to ${\cal N}=2$ super
Liouville field theory. In this and the next subsections, we restrict ourselves
to the bulk case for the simplicity of expressions, and in the last
subsection, the analysis is extended to the case with boundary.

The super coset is defined by the ${\cal N}=1$ supersymmetric
SL(2) WZNW model gauged by a $U(1)$ direction, where
the supersymmetry is enhanced to ${\cal N}=2$.
In addition to
the bosonic SL(2) currents $j^a ~(a=1,2,3)$ with level
$k_B = k + 2$,
we have three fermions $\psi^a$ with OPEs
$\psi^a (z) \psi^b (0) \sim \delta^{ab}/z$.
It is convenient to bosonize  as
\begin{align}
\psi^\pm = \frac{1}{\sqrt{2}} (\psi^1 \pm i \psi^2) ~, \qquad
\psi^\pm  = e^ { \pm i \sqrt{2} H_L } ~,
\end{align}
where $H_L(z) H_L(0) \sim - 1/2 \ln z$. The $U(1)$ direction we are gauging is
generated by
\begin{align}
 J^3 = j^3 + \psi^+ \psi^- = j^3 + i \sqrt{2} \partial H_L
     = i \sqrt{\frac{k}{2}} \partial X_L ~,
\end{align}
where the last equality defines one free boson $X_L$ with
$X_L(z) X_L(0) \sim - 1/2 \ln z$.
One of the fermions $\psi^3$ is also decoupled due to the gauging procedure.
Above we have discussed the holomorphic part, but the
anti-holomorphic part can be defined in the same way.

In total, the action is given by
\begin{align}
 S = S^H
 + \frac{1}{ \pi} \int d^2 w \partial H \bar \partial H
 + \frac{1}{ \pi} \int d^2 w \partial X \bar \partial X
 \, \left ( + S_{b,c} \right ) ~,
\end{align}
where we use the action $S^H$
for the $H_3^+$ model as in \eqref{Haction}.
Here we combined holomorphic and anti-holomorphic parts as
$H=H_L + H_R$ and $X=X_L+X_R$.
This action should be obtained through the standard procedure of
\cite{Gawedzki:1991yu} as well.
The ghost system with $b,c$ enters through the gauge fixing but
it will decouple from the other parts as before.
The parameter is now set as $Q_\phi = b = 1/\sqrt{k_B - 2} = 1/\sqrt{k}$.
Vertex operators invariant under the gauge transformation are
given by (see e.g. \cite{Giveon})
\begin{align}
 \Psi^{j,s}_{m , \bar m} (z)
  = \Phi^j_{m,\bar m} (z) V^H_{s, \bar s} (z)
 V^X_{m + s, \bar m + \bar s} (z) ~.
\end{align}
Here $\Phi^j_{m , \bar m}$ is a primary operator of the $H_3^+$ model,
which is  defined in \eqref{Hvertex}.
The other vertex operators are
\begin{align}
 V^H_{s, \bar s} (z)  = e^{i \sqrt{2} ( s H_L - \bar s H_R) } ~,
 \qquad
  V^X_{ m + s, \bar m + \bar s} (z)
 = e^{i \frac{2}{\sqrt{k}} ( ( m + s ) X_L - (\bar m + \bar s) X_R )} ~.
\end{align}
The correlation function is now written as
\begin{align} \label{sccorr}
& \left \langle \prod^N_{a = 1} \Psi^{j_a , s_a}_{m_a , \bar m_a } (z_a )
 \right \rangle
 = \prod^N_{a = 1}  \left [ N_{m_a  , \bar m_a }^{j_a }
  \int \frac{d^2 \mu_a }{| \mu_a |^2} \mu_a^{m_a}
 \bar \mu_a^{\bar m_a } \right ] \times \\
  & \qquad \qquad \times  \left \langle V^X_{- \frac{kS}{2} , - \frac{kS}{2}} (\zeta)
  V^H_{S , S} (\zeta)
 v^S (\zeta) \prod^N_{a = 1} V_{j_a} (\mu_a | z_a )
 V^H_{s_a, \bar s_a } (z_a ) V^X_{m_a + s_a , \bar m_a + \bar s_a } (z_a )
 \right \rangle ~. \nonumber
\end{align}
Here $S$ represents the violation  of total winding number
as $\sum_a (m_a +s_a ) = \sum_a (\bar  m_a + \bar  s_a) = kS/2$ and $v^S(\zeta)$ denotes the
spectral flow operator as before.
The operator $v^S(\zeta)$ again means that $\beta$ has a zero of order $S$ at $\zeta$
and the vertex operator $\exp (S \phi /b )$
is inserted.

\subsection{Duality with ${\cal N}=2$ super Liouville field theory}

Since the fermionic cigar model and ${\cal N}=2$ super Liouville theory
both preserve ${\cal N}=2$ super conformal symmetry, we just need to
show the correlators of tachyon vertex operators for the two
theories agree.
Moreover, in the correlation function \eqref{sccorr}, the fermion sector
with $H$ enters only through the direct products. It is thus natural
to expect that we can apply the same method as in \cite{HS,HS3} at least
to the SL(2) sub-sector. With the rotation of fields involving $H$, we
can show that the correlator in \eqref{sccorr} is mapped to the one
of the ${\cal N}=2$ super Liouville theory.

We follow the strategy in \cite{HS,HS3}, which was reviewed in the
previous section. We integrate out first $\gamma,\bar \gamma$ and
then $\beta, \bar \beta$. Then $\beta$ and $\bar \beta$ are
replaced by ${\cal B}$ and $ - \bar {\cal B}$ with
\begin{align}
 {\cal B} (w) =  \sum_{a = 1}^N \frac{\mu_\nu}{w - z_a}
   = u \frac{(w - \xi) ^S \prod_{a'=1}^{N-2-S} (w - y_{a'})}
 { \prod_{a = 1}^N (w - z_a)} ~.
\end{align}
In order to remove ${\cal B}$ from the action, we perform a
shift of $\phi$ as
\begin{align}
  \varphi (w,\bar w)
= \phi (w,\bar w) + \frac{1}{2b} \ln | {\cal B} (w) |^2 ~.
\end{align}
We furthermore perform a shift of $H_L$ along with $X_L$ as
\begin{align}
  \chi_L (w)  = X_L (w) - i \frac{\sqrt{k}}{2} \ln {\cal B} (w) ~, \qquad
  h_L (w) = H_L (w)  + i \frac{1}{\sqrt{2}} \ln {\cal B} (w) ~.
\end{align}
The anti-holomorphic parts $\chi_R$ and $h_R$ are defined by the complex
conjugates.
With the above new fields, the correlation function becomes
\begin{align}
 &\left \langle \prod^N_{a = 1} \Psi^{j_a , s_a}_{m_a , \bar m_a } (z_a)
 \right \rangle
 = \prod^{N-2-S}_{ i = 1} \int \frac{d^2 y_i}{(N-2-S)!}
  \prod^N_{a = 1}  N_{m_a , \bar m_a }^{j_a }
  \times \\
  & \times  \left \langle  \prod^N_{a = 1} V_{\alpha_a } ( z_a )
 V^h_{s_a + 1, \bar s_a + 1} (z_a )
  V^\chi_{m_a + s_a - \frac{k}{2}, \bar m_a + \bar s_a - \frac{k}{2}} (z_a)\prod^{N-2-S}_{a' = 1} V_{-\frac{1}{2b}} (y_{a'})
 V^h_{-1,-1} (y_{a'}) V^\chi_{\frac{k}{2}, \frac{k}{2}} (y_{a'})
 \right \rangle ~, \nonumber
\end{align}
where the right hand side is computed with the action
\begin{align} \label{fcaction}
 S = \frac{1}{\pi} \int d^2 w \left(
 \partial \varphi \bar \partial \varphi  +
 \partial h \bar \partial h+ \partial \chi \bar \partial \chi
 + \frac{\sqrt{g}}{4}
 {\cal R} (Q_\varphi \varphi
 + Q_{\tilde h} \tilde h + Q_{\tilde \chi} \tilde \chi) + \pi \lambda e^{2 b \varphi}  \right) ~.
\end{align}
Here $\tilde \chi $ and $\tilde h$ are the dual fields and
background charges are $Q_\varphi = b + b^{-1}$,
$Q_{\tilde \chi} = - i \sqrt{k}$ and $Q_{\tilde h} = i \sqrt{2}$.
The vertex operator is $V_\alpha = \exp (2 \alpha \varphi)$
with $\alpha = b(j + 1) + 1/2b$.
In this way we rewrite the correlation function \eqref{sccorr}
in terms of bosonic Liouville field theory with $\varphi$
and two additional free bosons with $\chi,h$.

As in the bosonic case we first apply the self-duality of the
Liouville field theory under $b \leftrightarrow b^{-1}$,
and we then treat the vertex operators inserted at
$y_{a'}$ as a perturbation operator.
Now we have the interactions as
\begin{align}
 {\cal L}_1 = \tilde \lambda e^{2 b^{-1} \varphi} ~, \qquad
 {\cal L}_2 = - e^{- b^{-1} \varphi + i \sqrt{k} \tilde \chi - i \sqrt{2} \tilde h} ~.
\end{align}
The dual parameter $\tilde \lambda$ is defined in \eqref{dpara}.
Next we look for suitable field redefinitions.
First we take a linear combination of $\chi$ and $h$ such
that a new field has no background charge.
An orthogonal basis is given by
\begin{align}
 \tilde \chi^+ = \frac{1}{\sqrt{k_B}} \left( \sqrt{2} \tilde \chi + \sqrt{k} \tilde h \right) ~, \qquad
 \tilde \chi^- = \frac{1}{\sqrt{k_B}} \left( \sqrt{k} \tilde \chi - \sqrt{2} \tilde h \right) ~,
 \label{chipm}
\end{align}
whose background charges are $Q_{\tilde \chi^+} = 0$ and
$Q_{\tilde \chi^-} = - i \sqrt{k_B}$.
With these new fields,
the interaction terms become
\begin{align}
 {\cal L}_1 = \tilde \lambda
e^{2 b^{-1} \varphi} ~, \qquad
 {\cal L}_2 = - e^{- b^{-1} \varphi + i \sqrt{k_B} \tilde \chi^-} ~.
\end{align}
Fortunately, these interactions are exactly the same as \eqref{bulkint}
in the
bosonic case. Therefore, the rest is almost the same as before.
We perform field redefinitions as
\begin{align}
 \phi = (k_B - 1) \varphi - i \sqrt{k_B} b^{-1} \tilde \chi^- ~,
 \qquad
 \tilde X^- = - i \sqrt{k_B} b^{-1} \varphi - (k_B-1) \tilde \chi^- ~,
 \label{midfd}
\end{align}
giving the background charges as $Q_\phi = b =1/\sqrt{k}$ and
$Q_{\tilde X} = 0$.
Then we have
\begin{align}
 {\cal L}_1 = - \tilde \lambda \pi^{-1 - 2 b^{-2}}  \gamma (1 + b^{ - 2} )   e^{b^{-1} \phi + i \sqrt{k_B} \tilde X^- } ~, \qquad
  {\cal L}_2 = - e^{b^{-1} \phi - i \sqrt{k_B} \tilde X^- } ~,
\end{align}
where we have utilized the reflection relation of Liouville theory
with $\hat b = i/\sqrt{2}$.

In order to obtain ${\cal N}=2$ super Liouville field theory,
we need to rotate the fields furthermore.
We consider
\begin{align}
 \tilde X = \frac{1}{\sqrt{k_B}} \left( \sqrt{k} \tilde X^- + \sqrt{2} \tilde \chi^+ \right) ~, \qquad
 \tilde H = \frac{1}{\sqrt{k_B}} \left( - \sqrt{2} \tilde X^- + \sqrt{k} \tilde \chi^+ \right) ~,
 \label{XH}
\end{align}
such that the field content is the same as that for ${\cal N}=2$ super
Liouville field theory.
Namely, we have $\phi$ with
background charge $Q_\phi = b = 1/\sqrt{k}$, a free boson $X$
and a bosonized fermion $H$.
See also appendix C of \cite{Maldacena}.
The interaction terms are changed to
\begin{align}
 {\cal L}_1 = - \tilde \lambda \pi^{-1 - 2 b^{-2}}  \gamma (1 + b^{ - 2} )  e^{\sqrt{k}  ( \phi + i \tilde X ) - i \sqrt{2} \tilde H} ~, \qquad
  {\cal L}_2 = - e^{\sqrt{k} ( \phi - i \tilde X ) + i \sqrt{2} \tilde H}
  ~,
\end{align}
which are those for ${\cal N}=2$ super Liouville field theory.
The coefficients of the interaction terms
can be changed by the shift of zero mode of $\tilde X$ as
\begin{align}
 {\cal L}_1 = \tilde c
 e^{\sqrt{k}  ( \phi + i \tilde X ) - i \sqrt{2} \tilde H}~, \qquad
  {\cal L}_2 = \tilde c e^{\sqrt{k} ( \phi - i \tilde X ) + i \sqrt{2} \tilde H} ~,
\end{align}
where $\tilde c$ is as in \eqref{tildec}.

Moving to the vertex operators, we rewrite them
in a suitable form as
\begin{align}
 &V_{\alpha} ( z )
 V^H_{s + 1, \bar s + 1} (z)
  V^X_{m + s - \frac{k}{2}, \bar m + \bar s - \frac{k}{2}} (z)
 \\ & \nonumber =
  e^{2b(j+1+\frac{1}{2b^2})\phi +
 i \frac{2}{\sqrt{k_B}} \left( (m - \frac{k_B}{2}) \chi^-_L -
(\bar m - \frac{k_B}{2}) \chi^-_R \right)  } \cdot
e^{2 i \sqrt{\frac{2}{k k_B}} \left( (\frac{k_B}{2} s  + m) \chi^+_L
 - (\frac{k_B}{2} \bar s  + \bar m) \chi^+_R  \right)}
\end{align}
in terms of the new fields \eqref{chipm}.
Then we observe that the first factor on the right hand side
is of the same form as in the bosonic case.
Therefore, we can perform the reflection relation in
the same way. The result is
\begin{align}
& N^j_{m , \bar m} V_{\alpha} ( z )
 V^H_{s + 1, \bar s + 1} (z)
  V^X_{m + s - \frac{k}{2}, \bar m + \bar s - \frac{k}{2}} (z) \\
  & \qquad \qquad
 = -  \pi^{-1-2j-m-\bar m}  e^{\frac{2}{\sqrt{k}} \left ( (j+1) \phi
 +  i (m + s) X_L - i (\bar m + \bar s) X_R \right )
 + i \sqrt{2} (s H_L - \bar s H_R )  } ~, \nonumber
\end{align}
which are the vertex operators of ${\cal N}=2$ super Liouville field
theory.
In the last equation we have used the field redefinition \eqref{XH}.
Thus we have established the relation between correlation functions as
\begin{align}
 \left \langle \prod^N_{a = 1} \Psi^{j_a , s_a}_{m_a , \bar m_a} (z_a)
 \right \rangle  ^\text{fcig}
 = {\cal N}
  \left \langle \prod^N_{a = 1}  e^{\frac{2}{\sqrt{k}} \left ( (j_a+1) \phi
 +  i (m_a + s_a) X_L - i (\bar m_a + \bar s_a) X_R \right )
 + i \sqrt{2} (s_a H_L - \bar s_a H_R )  } (z_a)
 \right \rangle ^\text{N=2L} ~, \nonumber
\end{align}
where the left hand side and the right hand side are respectively computed
in the fermionic cigar model and in the ${\cal N}=2$ super
Liouville field theory.

\subsection{Fermionic FZZ duality with boundary}
\label{bffzz}

The above analysis shows that the fermionic FZZ duality naturally
comes from the bosonic FZZ duality through the proper change of fields.
In particular, it is easy to extend it to the case with closed Riemann
surfaces of arbitrary genus. In this subsection
the fermionic FZZ duality is generalized to disk amplitudes with
proper boundary conditions.
We consider D1-branes in the fermionic 2d black hole, which
should be mirror to B-branes in the ${\cal N}=2$ super Liouville
field theory studied in \cite{Hosomichi}.
Since the analysis is almost the same as before, we
explain it only briefly.

The boundary action for D1-branes is the same as the one for the bosonic case
in \eqref{bint}
\begin{align}
 S_\text{bint} = i \lambda_B \sigma_3 \int d u \beta e^{b \phi} ~,
 \qquad
 \lambda_B = \sqrt{  \frac{ \lambda}{\sin \pi b^2}} \sinh r ~,
\end{align}
but with the bulk action in \eqref{fcaction}.
We assign Dirichlet boundary conditions as $\beta + \bar \beta = 0$,
$\gamma + \bar \gamma = 0$, and Neumann boundary conditions for $\phi$.
We also assign $X_L + X_R = \sqrt{k} \theta_0$ and
$\psi^\pm = e^{\pm 2 i \alpha} \bar \psi^\mp$ for the fermions.
Boundary operators are defined as
\begin{align}
 \Psi^{l,s}_{m,i} (z)
  = \frac{\sigma_i}{\sqrt2}  \Phi^l_{m , \eta} V^H_{s} V^X_{m+s} (z) ~,
\end{align}
where the operator for $H_3^+$ is given in \eqref{bo} and the other
operators are
\begin{align}
 \qquad
  V^H_{ s } (z)
 = e^{i \sqrt{2} s H_L } ~, \qquad
   V^X_{ m + s } (z)
 = e^{i \frac{2}{\sqrt{k}} ( m + s ) X_L} ~.
\end{align}
Since the boundary operator changes the boundary conditions generically,
we may label as $[\Psi^{l,s}_{m,i}]_{L,L'}$ with $L = (r,M,\alpha)$.
The labels take values $s = \alpha - \alpha ' + S$
and $m + s = M - M ' + \alpha - \alpha ' + \frac{k}{2} w$ with $S,w \in \mathbb{Z}$.
We should use $w \in 2 \mathbb{Z}$ for $i=0,3$ and
$w \in 2 \mathbb{Z} + 1$ for $i=1,2$ as before.
Similarly, $\eta = 0$ for $i=0,1$ and  $\eta = 1$ for $i=2,3$.

Once we have the boundary operators, the rest is straightforward.
However, to compare with known results, let us first discuss the Chan-Paton factors.
It is possible to rewrite the non-Abelian action
by an Abelian action with the introduction of boundary fermions,
see \cite{AT,Tseytlin,KL,TTU}. One good review is in section 3 of \cite{KL}.
Basically, when we compute correlation functions with an action including matrix coefficients, we have to perform a trace with keeping
path ordering. The same effect can be made with the introduction of
boundary fermions $\Theta, \bar \Theta$. The kinetic term may be
given by
\begin{align}
 S = \int du \Theta \partial_u \bar \Theta ~,
\end{align}
which leads to
\begin{align}
 \langle \Theta (u_1) \bar \Theta (u_2) \rangle =
  \langle \bar \Theta (u_1) \Theta (u_2) \rangle = \frac{1}{2}
 \text{sgn} (u_1 - u_2) ~.
\end{align}
Then the algebra of $\sigma_+ , \sigma_- , \sigma_3$ can be reproduced
by $\sqrt2 \Theta , \sqrt2 \bar \Theta ,
2 (\Theta \bar \Theta - \bar \Theta \Theta ) $.
However, the off-diagonal terms become Grassmann odd in this formulation.

With the above argument in mind, we proceed further.
Following the previous analysis we find the boundary interaction terms as
\begin{align}
 &{\cal L}_{B,1} =  i \sqrt{\frac{\tilde c}{2} }  \left( e^{\frac{r}{2b^2}} \sigma_+
 + e^{- \frac{r}{2b^2}} \sigma_-  \right)
 e^{\frac{1}{2 b}\phi + i \frac{\sqrt{k_B}}{2} \tilde X^-  }  ~, \\
 &{\cal L}_{B,2} =  i \sqrt{\frac{\tilde c}{2}}
 \left(e^{- \frac{r}{2b^2}} \sigma_+ + e^{\frac{r}{2b^2}} \sigma_- \right)
 e^{\frac{1}{2 b}\phi - i \frac{\sqrt{k_B}}{2} \tilde X^-  }  \nonumber
\end{align}
in terms of the fields \eqref{midfd}.
Applying the rotation in \eqref{XH} and rewriting the Chan-Paton factors,
we now have%
\footnote{This form of boundary interactions suggests that we are dealing
with a $D \bar D$ system. Moreover, open string tachyons between brane
and anti-brane are condensed in this system. See \cite{KL,TTU} in more details.}%
\begin{align}
 &{\cal L}_{B,1} =  i \sqrt{\tilde c }  \left( e^{\frac{r}{2b^2}} \Theta
 + e^{- \frac{r}{2b^2}} \bar \Theta  \right) \psi^-
e^{\frac{\sqrt{k}}{2} (\phi + i \tilde X) }   ~, \\
 &{\cal L}_{B,2} =  i \sqrt{\tilde c}
 \left(e^{- \frac{r}{2b^2}} \Theta + e^{\frac{r}{2b^2}} \bar \Theta \right)\psi^+
 e^{\frac{\sqrt{k}}{2} (\phi - i \tilde X) }  ~. \nonumber
\end{align}
One advantage to introduce the boundary fermions is on the co-cycle factor.
Since we have fermionized as $\psi^\pm = e^{\pm i \sqrt{2} H_L}$, we should
take care of its Grassmann parity. Fortunately, we replaced the
Pauli matrices $\sigma^\pm$ by boundary fermions at the same time, thus
the boundary interaction terms remain bosonic.
Notice the above interaction terms are the same as (5.8) with (5.27) of
\cite{Hosomichi}.
In particular, matrix factorization suggests that the bulk cosmological
constant is the square of boundary cosmological constant.

We should work out the vertex operators as well.
With the same method as before, we find that the vertex operators are
mapped to
\begin{align}
 B^{l,s}_{m,i} = \frac{\sigma_i}{\sqrt2}
e^{\frac{1}{\sqrt{k}} \left ( (l+1) \phi
 +  i (m + s) (X_L-X_R) \right )
 + i \sqrt{2} s H_L   } ~.
\end{align}
We should again replace the Chan-Paton factor by boundary fermions.
Then, these boundary operators coincide with those in \cite{Hosomichi}
(see eq.(5.18) of the paper and arguments given below).
Therefore, we have reproduced the results for B-branes in ${\cal N}=2$
super Liouville field theory from D1-branes in the fermionic
2d black hole.

\section{Outlook}

In this article, we derive  the FZZ duality \cite{FZZ2} for disc correlators,
which is a strong/weak duality between the cigar model and
sine-Liouville field theory.
For the purpose, we have used perturbative path integral methods of \cite{HS3} in combination with the strong/weak self-duality of Liouville field theory.
 This was extended to the duality between the supersymmetric cigar and $\mathcal N=2$ super-Liouville theory as in
\cite{HK} both for correlators on the sphere and the disc.
There are several possible applications, generalizations and future directions.

For D1- or A-branes in $\mathcal N=2$ super Liouville theory,
a Lagrangian description is known \cite{Hosomichi}. We would expect these branes to be dual to branes in the cigar which wind the circular direction. Thus for D2-branes in the (super)cigar model, it might be possible to find a boundary Lagrangian and derive the duality with the branes in sine-Liouville (super Liouville) theory.
D0-branes in sine-Liouville theory are not covered by our derivation, so one should study these branes directly in the sine-Liouville theory.
They are of ZZ-type \cite{ZZ} and thus can be studied by looking at Cardy conditions.

One should be able to apply our methods to other models.
One example is the coset OSP(1$|$2)/$\mathbb R$ studied in \cite{Giribet:2009eb}.
The OSP(1$|$2) WZNW model is in correspondence with $\mathcal N=1$ super Liouville field theory, and the derivation is analogous to the one between $H_3^+$ model and Liouville field theory
\cite{Hikida:2007sz, Creutzig:2010zp}.
It is likely that also the OSP(1$|$2)/$\mathbb R$ coset possesses a strong/weak dual which is yet to be determined. As our derivation is quite constructive, it might be possible to use it to find and derive the duality in one step.
The $H_3^+$-Liouville correspondence extends  to e.g. correspondences between OSP(N$|$2) and SU(M$|$2) supergroup WZNW models and  superconformal field theories respectively with SO(N)- and U(M)-extended supersymmetry \cite{CHR}. These extended supersymmetry algebras were introduced in \cite{Bershadsky:1986ms,Knizhnik:1986wc}. The derivation of these correspondence is again via the path integral and these superconformal field theories seem to be strong/weak self-dual.\footnote{Except for $\mathcal N=2$ super Liouville theory, which
is the case in this article.}
Again one can investigate cosets thereof \cite{Creutzig:2009fh} and look for possible weakly coupled dual theories.
This should be particularly interesting for cosets of the PSU(2$|$2) and OSP(4$|$2) WZNW models as these supergroups are important in the AdS/CFT correspondence.

\acknowledgments

We would like to thank L.~Dolan and K.~Hosomichi for useful discussions.
The work of YH is supported in part by Keio Gijuku Academic Development Funds, and the work of TC partially by U.S. Department of Energy,
Grant No. DE-FG02-06ER-4141801, Task A.

\appendix

\section{Branes}

In this appendix we recall geometry and action of branes in the coset.
Geometry of branes in cosets has been considered in \cite{Gawedzki:2001ye,Fredenhagen:2001kw} and the boundary action of AdS$_2$ branes was given in \cite{FR}.

\subsection{Geometry of Branes}\label{app:geometry}

An element in $H_3^+$ satisfies $g^\dagger=g$. $AdS_2$ branes are described by the restriction of twisted conjugacy classes of $SL(2,\mathbb C)$ to $H_3^+$, i.e.
\begin{equation}
C_a^\omega \ = \ \{ hah^t \ | \ h\,\in\, SL(2,\mathbb{R})\ \} \, .
\end{equation}
Here, $a=a^\dagger$ is in $H_3^+$ and hence the same is true for every element of $C_a^\omega$.
The coset $H_3^+/\mathbb R$ is the twisted adjoint one, consisting of elements $g$ in $H_3^+$ modulo the twisted adjoint action
\begin{equation}\label{eq:twistedadj}
g \ \sim \ ugu
\end{equation}
where $u=u^t=u^\dagger$ in the $t^3$ direction.
Branes in cosets are now described by
elements of the form $gv^{-1}$ where $g$ is in the twisted conjugacy class describing the $H_3^+$ brane and $v$ in the conjugacy class for
the subgroup we mod out, i.e. $\mathbb R$ \cite{Fredenhagen:2001kw}.
In addition, this translate of a twisted conjugacy class has to be invariant under the twisted adjoint action defining the coset \eqref{eq:twistedadj}.
For us this is true for either Dirichlet ($v$ constant) or Neumann boundary conditions in the $\mathbb R$ directions, i.e.
\begin{equation}
uhah^tv^{-1}u \ = \ uhah^tuv^{-1}\ = \ uhah^tu^tv^{-1}\ = \ uha(uh)^tv^{-1} \, .
\end{equation}
We choose to consider Dirichlet boundary conditions, i.e. $v = e^{i\theta_0} =$ constant.
The embedding of the gauged WZNW model in the product theory of $H_3^+\times U(1)$ is realized by parameterizing the gauge fields as $A=i\del X$, $\bar A=-i\bar\del X$ and changing variables $g'\ = \ e^{iX t^ 3}ge^{-iX t^ 3}$.

\subsection{Boundary actions and Chan-Paton factors}\label{app:chanpaton}

In this subsection, we argue for consistent boundary actions including boundary fermions, which can be seen as
Chan-Paton factors.
The bulk action is
\begin{equation}
S_{\text{bulk}}\ = \ \frac{1}{2\pi} \int d^2z\, \bigl(\del\phi\bar\del\phi-\gamma\bar\del\beta-\bar\gamma\del\bar\beta-\mu b^2\beta\bar\beta e^{2b\phi}\bigr)\, .
\end{equation}
As a boundary term, we choose
\begin{equation}
S_{\text{bdy}}\ = \ \frac{i\mu_B}{2\pi} \int du\, \bigl(\lambda\del_u\bar\lambda+f(\lambda, \bar\lambda)\beta e^{b\phi}\bigr)\, ,
\end{equation}
and we vary under the Dirichlet constraint $\beta=-e^{i\theta_0}\bar\beta$.  We want to determine functions $f(\lambda,\bar\lambda)$ that are consistent with preserving current symmetry at the boundary.
Then the variation of the action has the following boundary contribution
\begin{equation}
\begin{split}\label{eq:bdyvariation}
\delta S\bigl|_{\text{bdy}} \ = \  \frac{i}{2\pi} \int du\, \bigl(& \delta\phi((\bar\del-\del)\phi +f(\lambda,\bar\lambda)bc\beta e^{b\phi})+
           \delta\beta(\gamma+e^{-i\theta_0}\bar\gamma+ \\ &+f(\lambda,\bar\lambda)\mu_B e^{b\phi})+
	   \delta\lambda (\del_u\bar\lambda+ \frac{d}{d\lambda} f(\lambda, \bar\lambda) \mu_B\beta e^{b\phi})+ \\
	   &+\delta\bar\lambda (\del_u\lambda+ \frac{d}{d\bar\lambda} f(\lambda, \bar\lambda) \mu_B\beta e^{b\phi})\bigr)\, .
\end{split}
\end{equation}
Moreover, the bulk equations of motion imply
\begin{equation}\label{eq:eombeta}
\mu b^2 e^{2b\phi}\bar\beta\ = \ \bar\del\gamma \qquad\text{and}\qquad \mu b^2 e^{2b\phi}\beta\ = \ \del\bar\gamma\, .
\end{equation}
The currents take the following form
\begin{equation}
\begin{split}
 J^- \ &= \ \beta \ \ \ , \ \ \
 J^3 \ = \ \beta\gamma+b^{-1}\del\phi \ \ \ , \ \ \
 J^+ \ = \ \beta\gamma^2+2b^{-1}\gamma\del\phi-(k-2)\del\gamma \, , \\
 \bar J^- \ &= \ \bar\beta \ \ \ , \ \ \
 \bar J^3 \ = \ \bar\beta\bar\gamma+b^{-1}\bar\del\phi \ \ \ , \ \ \
 \bar J^+ \ = \ \bar\beta\bar\gamma^2+2b^{-1}\bar\gamma\bar\del\phi-(k-2)\bar\del\bar\gamma\, . \\
\end{split}
\end{equation}
The gluing conditions for AdS$_2$ branes are
\begin{equation}
\begin{split}
J^- \ &= -\ e^{i\theta_0}\bar J^- \ \ \ , \ \ \
J^3 \ = \ \bar J^3 \ \ \ \text{and} \ \ \
J^+ \ = -\ e^{-i\theta_0}\bar J^+ \, .
\end{split}
\end{equation}
The vanishing of the variation of above boundary part of the action implies that classically $J^-=-e^{i\theta_0}\bar J^-$ and $J^3=\bar J^3$.
It also implies that $J^+ = -e^{-i\theta_0}\bar J^+$ as follows.
Using $J^-=-e^{i\theta_0}\bar J^-$ and $J^3=\bar J^3$, we see that the conditions
\begin{equation}
 \begin{split}\label{eq:Jplus}
J^+ \ &= -\ e^{-i\theta_0}\bar J^+ \qquad \Leftrightarrow\\
0\ &= \ -b^{-1}(e^{-i\theta_0}\bar\gamma+\gamma)(\del+\bar\del)\phi +(k-2)e^{-i\theta_0}\bar\del\bar\gamma+(k-2)\del\gamma \\
 \end{split}
\end{equation}
are equivalent. Using the bulk equations of motion \eqref{eq:eombeta}, $\beta=-e^{i\theta_0}\bar\beta$ and $\delta S =0$ \eqref{eq:bdyvariation}, we rewrite
\begin{equation}
 \begin{split}
  \del\gamma+e^{-i\theta_0}\bar\del\bar\gamma \ &= \ (\del+\bar\del)(\gamma+e^{-i\theta_0}\bar\gamma) +e^{-i\theta_0}\del\bar\gamma-\bar\del\gamma \\
&= \  (\del+\bar\del)(\gamma+e^{-i\theta_0}\bar\gamma)\qquad \text{because of}\ \eqref{eq:eombeta} \ \text{and} \ \beta\ = \ -e^{i\theta_0}\bar\beta \\
&= \ -(\del+\bar\del) \bigl(f(\lambda, \bar\lambda)\mu_Be^{b\phi}\bigr) \qquad \text{because of}\ \eqref{eq:bdyvariation} \,.
 \end{split}
\end{equation}
Inserting this in \eqref{eq:Jplus} and using $b=1/\sqrt{k-2}$, we see that
classically
\begin{equation}
 \begin{split} 
J^+ \ &= \ - e^{-i\theta_0}\bar J^+ \qquad \Leftrightarrow\qquad
0\ = \ (\del+\bar\del)f(\lambda,\bar\lambda)\, . \\
 \end{split}
\end{equation}
Recall the equations of motion of $\lambda$ and $\bar\lambda$ from \eqref{eq:bdyvariation}
\begin{equation}
0 \ = \ \del_u\bar\lambda+ \frac{d}{d\lambda} f(\lambda, \bar\lambda) \mu_B\beta e^{b\phi}\ = \
	   \del_u\lambda+ \frac{d}{d\bar\lambda} f(\lambda, \bar\lambda) \mu_B\beta e^{b\phi} \, .
\end{equation}
Hence, $(\del+\bar\del)f(\lambda, \bar\lambda)$ vanishes if it is one of the following
\begin{equation}
\begin{split}
f(\lambda, \bar\lambda)\ &= \ c \, ,\\
f(\lambda, \bar\lambda)\ &= \ c\lambda \, ,\\
f(\lambda, \bar\lambda)\ &= \ c\bar\lambda \, ,\\
f(\lambda, \bar\lambda)\ &= \ c\lambda\bar\lambda \, \\
\end{split}
\end{equation}
for some constant $c$. This can be related to the u(2) algebra and  means that the Chan-Paton factors $\sigma_0$, $\sigma_3$ and $\sigma_\pm=\frac12 (\sigma_1\pm i\sigma_2)$ are consistent with preserving current symmetry.

\section{The gauged sigma model}
\label{app:gaugedsigma}

\subsection{The bulk theory}

To warm up we consider the coset $H_3^+/\mathbb R$ on the sphere.

\subsubsection*{Coset as gauged sigma model}

As in \cite{HS3} we write elements in $H_3^+$ as
\begin{equation}\label{}
    h=e^{\gamma \sigma^+}e^{\phi \sigma^3}e^{\bar\gamma \sigma^-}\ .
\end{equation}
The symmetries are $g\mapsto B h B^\dagger$ generated by $aJ-a^*\bar J$ where we use\footnote{Note that the components for the anti-chiral currents in last appendix are related to this definition with an extra minus transposed automorphism.}
\begin{equation}\label{}
    J=-k\del h h^{-1},\qquad \bar J = k h^{-1}\bar\del h.
\end{equation}
We gauge the direction corresponding to $J_3-\bar J_3=\tr (J-\bar J)\sigma^3$,
the gauged model is
\begin{equation}\label{eq:gaugedaction}
    S=\frac{k}{2\pi}\int\,d^2z\left((\bar\del\phi+\bar A)(\del\phi+A)+e^{-2\phi}(\bar\del+\bar A)\gamma(\del+A)\bar\gamma\right)\ .
\end{equation}
This is invariant under the gauge symmetry
\begin{equation}\label{eq:gaugetrans}
    h\mapsto e^{\lambda(z,\bar z)\sigma^3/2}h e^{\lambda(z,\bar z)\sigma^3/2},\qquad A\mapsto A-d\lambda(z,\bar z),
\end{equation}
where we have used the symbol $A$ for the whole one-form as well as its holomorphic component.
On the fields this acts as
\begin{equation}\label{}
    \phi\mapsto\phi+\lambda(z,\bar z),\qquad \gamma\mapsto e^{\lambda(z,\bar z)}\gamma,\qquad \bar \gamma\mapsto e^{\lambda(z,\bar z)}\bar\gamma\ .
\end{equation}

To get the cigar model we note that the equations of motion for $A,\bar A$ are
\begin{align}\label{eq:eqmA}
    A&=-\frac{\del\phi+e^{-2\phi}\gamma\del\bar\gamma}{1+e^{-2\phi}\gamma\bar\gamma}=-\del\phi-\frac{v\del\bar v}{1+v\bar v}\ ,\\
    \bar A&=-\frac{\delbar\phi+e^{-2\phi}\bar\gamma\delbar\gamma}{1+e^{-2\phi}\gamma\bar\gamma}=-\delbar\phi-\frac{\bar v\delbar v}{1+v\bar v}\ ,
\end{align}
where we have introduced the gauge invariant coordinates
\begin{align}\label{eq:defivcoordinates}
    v=e^{-\phi}\gamma\ ,\qquad \bar v=e^{-\phi}\bar\gamma\ .
\end{align}
The action then takes the cigar form
\begin{align}\label{}
    S=\frac{k}{2\pi}\int\,d^2z\left(\frac{1}{1+v\bar v}\bar\del v\del\bar v\right)\ .
\end{align}

On the other hand we can fix the gauge field using a complex valued gauge $U=\exp(\alpha+ix)$, note that $x$ is here $2\pi$ periodic,
\begin{equation}\label{}
    A=U^{-1}\del U=\del \alpha+i\del x,\qquad \bar A=(A)^*=U^{*-1}\bar\del U^*=\bar\del \alpha-i\bar\del x.
\end{equation}
The real part, $\alpha$, is pure gauge and the integration over $x$ gives the volume of the gauge group, but the imaginary part cannot be gauged away, but can be decoupled by making a transformation conjugate to the gauge symmetry i.e. along $i J_3+i\bar J_3$:
\begin{equation}\label{}
    h\mapsto e^{i x(z,\bar z)\sigma^3/2}h e^{-ix(z,\bar z)\sigma^3/2}
\end{equation}
which simply takes
\begin{equation}\label{}
\gamma\mapsto e^{ix(z,\bar z)}\gamma,\qquad \bar\gamma\mapsto e^{-ix(z,\bar z)}\bar\gamma \ .
\end{equation}
This gives the action of the product theory $H_3^+\times$U(1)
\begin{equation}\label{}
    S=\frac{k}{2\pi}\int\,d^2z\left(\bar\del\phi\del\phi+\bar\del x\del x+e^{-2\phi}\bar\del\gamma\del\bar\gamma\right)\ .
\end{equation}

\subsubsection*{Strings without winding}

Let us now consider what happens to the vertex operators. The primary fields of $H_3^+$ in the $m$-basis may be written (after a rescaling of $\phi$)
\begin{equation}\label{}
    \Phi^j_{m\bar m}\propto \gamma^{m-(j+1)}\bar\gamma^{\bar m-(j+1)}e^{2b(j+1)\phi},
\end{equation}
where
\begin{equation}\label{}
    m=\frac{n+ip}{2},\qquad\bar m=\frac{-n+ip}{2},\qquad n\in\mathbb{Z},p\in\mathbb{R}\ .
\end{equation}
Note that these are operators with no winding around the boundary of $H_3^+$.
We must require gauge invariance of our operators. The basis is such that $J_3=-2m,\bar J_3=2\bar m$. Thus we require $m+\bar m=i p=0$. Under the decoupling of the imaginary gauge field we then get
\begin{equation}\label{}
    \Phi^j_{m\bar m}\mapsto \Phi^j_{m\bar m} e^{(m-\bar m) i x}=\Phi^j_{m\bar m} e^{n i x},
\end{equation}
i.e. also no winding in the U(1) direction.
If we introduce currents for the U(1) part
\begin{equation}\label{}
    J_0=-ik\del x,\qquad \bar J_0=ik\bar\del x,
\end{equation}
we see that the total currents
\begin{equation}\label{eq:jtotal}
    J_3^{\mathrm{Total}}=J_3-2J_0,\qquad \bar J_3^{\mathrm{Total}}=(J_3^{\mathrm{Total}})^*=\bar J_3+2\bar J_0
\end{equation}
are zero when acting on the primary fields.

\subsubsection*{Winding strings}

We now consider states that wind asymptotically in $H_3^+$ i.e. spectrally flowed states.
To this end we define the scalar field $H$ by $J^3=2\del H$. If we use the first order formalism for $H_3^+$ and bosonize the $\beta,\gamma$ system such that $\gamma=\exp(Y_L-Z_L)$, $\beta=-\exp(-Y_L+Z_L)\del Z$ we have
\begin{equation}\label{eq:defiH}
    H=-\phi/b-\beta\gamma=-\phi/b-Y.
\end{equation}
This field has OPE $HH\sim-k/2\ln|z-w|^2$ and background charge $1/2$. The $w$ times spectrally flowed state (i.e. $w$ times winding) is then
\begin{equation}\label{}
    \Phi^{j w}_{m\bar m}=\,:e^{w H}\Phi^j_{m\bar m}:
\end{equation}
Under gauge transformations \eqref{eq:gaugetrans} we have $Y_L\mapsto Y_L+\lambda(z,\bar z)$, $Y_R\mapsto Y_R+\lambda(z,\bar z)$ and hence
\begin{align}\label{}
    H\mapsto H-k\lambda(z,\bar z)\ .
\end{align}
Thus demanding gauge invariance of these operators means
\begin{equation}\label{}
    m=\frac{n+kw}{2},\qquad\bar m=\frac{-n+kw}{2},\qquad n,w\in\mathbb{Z},
\end{equation}
i.e. $ip\mapsto w$. $H$ is invariant under the transformation decoupling the U(1) part so
\begin{equation}\label{eq:bulkoperator}
    \Phi^{j w}_{m\bar m}\mapsto\Phi^{j w}_{m\bar m}e^{in x}\equiv \Phi^{j w}_{m\bar m}V_n.
\end{equation}
So now we have winding in the $H_3^+$ part, but none in the U(1) part. Again the total currents \eqref{eq:jtotal} are zero.
We can use the spectral flow covariance of the correlators to move all spectral flow to the identity operator at infinity \cite{Ribault}. There will however be a prefactor on the correlator depending on the positions of the vertex operators. To avoid this we do a similar, but chiral spectral flow in the U(1) theory such that
\begin{equation}\label{eq:windinginu1}
    \Phi^{j w}_{m\bar m}V_n\mapsto \Phi^{j}_{m\bar m}V_{n w}\propto \gamma^{m-(j+1)}\bar\gamma^{\bar m-(j+1)}e^{2b(j+1)\phi}e^{i n x+i w k \tilde x},
\end{equation}
with $\tilde x=x_L-x_R$. Doing the spectral flow in both $H_3^+$ and U(1) also means that the total currents \eqref{eq:jtotal} are kept zero. Further, the dimension of the operators on the left and right hand side can be shown to be the same, and this is the reason that we get an equality of the correlators. For the identity operator at infinity in the case of violated winding number we have
\begin{equation}\label{eq:identiope}
    1\mapsto e^{-\sum_i w_i H}e^{-i\sum_i w_i k \tilde x}.
\end{equation}
Identifying $X=x/\sqrt{k}$, we have now reached the operators used in the main text which have no winding in $H_3^+$ direction, but in U(1).

\subsubsection*{Winding by Wilson lines}

As shown in \cite{Gawedzki:1991yu} (see eq. (50) in section 4) there is an alternative way to make $H^+_3$ vertex operators with $m+\bar m\neq0$ gauge invariant -- simply add Wilson lines:
\begin{equation}\label{eq:wilsonlinestate}
    \prod_i\Phi^{j_i }_{m_i\bar m_i}(\xi_i)e^{\int_C A},
\end{equation}
where $C$ is a chain with $\delta C=\sum_i(m_i+\bar m_i)\xi_i$. This is nicely gauge invariant. The demand that $m+\bar m$ is real in this case comes from locality.

When one decouples the gauge field, we exactly get the states \eqref{eq:windinginu1} which wind in the U(1) direction. To show that \eqref{eq:wilsonlinestate} and \eqref{eq:bulkoperator} gives the same state when going to the cigar model, one could try to construct the first order formalism for the gauged sigma model, but we will refrain from doing that here.

\subsection{The boundary theory}

We now consider the gauged model where the world-sheet is a disk. We are going to argue that the boundary action
\begin{align}
 S_\text{bint} = i \lambda_B e^{i\theta_0} \sigma_3 \int d u \beta e^{b \phi}
\end{align}
arises from a single brane in the cigar. The boundary conditions used for the decoupled $H_3^+$ model are ($c$ and $\lambda_B$ are proportional)
\begin{align}\label{eq:bdryinh3plus}
   e^{i\theta_0} \del\bar\gamma+e^{-i\theta_0}\bar\del\gamma&=0, \qquad (e^{i\theta_0}\beta+e^{-i\theta_0}\bar\beta=0),\nonumber \\ e^{-i\theta_0}\gamma+e^{i\theta_0}\bar\gamma&=\pm c e^{\phi},\qquad \delbar\phi-\del\phi=\pm c b e^{i\theta_0}\beta e^{b\phi},\qquad z=\bar z\ .
\end{align}

The boundary action, seen as a perturbative interaction term, will geometrically deform the brane with $c=0$ to the brane with non-zero $c$.

Let us start with a single brane in the cigar given by the boundary conditions
\begin{align}\label{eq:boundarycondcigar}
    e^{-i\theta_c}v+e^{i\theta_c}\bar v=2 Re(e^{-i\theta_c}v) &=c,
\end{align}
where the coordinates $v,\bar v$ were defined in \eqref{eq:defivcoordinates}. Of course, there is rotational invariance, so we could simply set $\theta_c=0$. Note that for $v=\sinh\rho e^{i(\theta-\pi/2)}$ we get
\begin{align}\label{}
    \sinh\rho \sin(\theta-\theta_c)=c/2 \propto\sinh r\ ,
\end{align}
as in \eqref{eq:braneeq}.

The boundary conditions in the gauged WZNW model would have to be a gauge invariant version of the ones in $H_3^+$ model. To descend to \eqref{eq:boundarycondcigar} they will be
\begin{multline}\label{eq:bdrygauged}
    e^{i\theta_c}(\del+A)\bar\gamma+e^{-i\theta_c}(\bar\del+\bar A)\gamma=0,\qquad e^{-i\theta_c}\gamma+e^{i\theta_c}\bar\gamma=c e^{\phi},\qquad \del\phi+A-\bar\del\phi-\bar A=c b e^{i\theta_c}\beta e^{b\phi},\\ \qquad z=\bar z.
\end{multline}
Further, one needs to assign boundary conditions to $A$. There are two obvious choices $A=\pm \bar A$. The boundary conditions above are gauge invariant by construction, so by gauge-fixing $A=d\alpha+i*d x$ we can remove $\alpha$. However, we cannot separate $x$ out of the condition unless we choose Dirichlet boundary condition on $x$ i.e.
\begin{equation}\label{}
    \del x+\bar \del x=0,\qquad A-\bar A=0\ ,
\end{equation}
which will be our choice of boundary conditions.

Going to the product theory by decoupling $x$ we get
\begin{align}\label{eq:derivedboundarycond}
    e^{-i (x_0-\theta_c)}\del\bar\gamma+e^{i (x_0-\theta_c)}\bar\del\gamma&=0\ , \qquad (e^{-i (x_0-\theta_c)}\beta+e^{i (x_0-\theta_c)}\bar\beta=0)\ ,\nonumber \\
     e^{i (x_0-\theta_c)}\gamma+e^{-i (x_0-\theta_c)}\bar\gamma&=c e^{\phi},\qquad \del\phi-\bar\del\phi=c b e^{-i(x_0-\theta_c)}\beta e^{b\phi},\qquad z=\bar z\ .
\end{align}
This is just the boundary condition for the rotated brane. Here $x_0$ is the boundary value of $x$
\begin{equation}\label{}
    x_L+x_R=x_0,\qquad z=\bar z\ .
\end{equation}

Before continuing let us consider the boundary states.  First, we should remove all momentum in the circular direction, i.e. set $n=0$ since our strings are attached to the D-branes. This also easily follows using the boundary conditions on the bulk states. With winding we expect two types of strings in the cigar. Strings with integer winding and strings with half-integer winding of course ending on the same brane since we only have one brane in the cigar, see figure \ref{fig:windingstring} (see also figures in \cite{RS}).

A boundary operator analogous to the bulk state \eqref{eq:windinginu1} is
\begin{equation}\label{}
    \Phi^{j}_{w}V_{ w}\propto \gamma^{kw/4-(j+1)/2}\bar\gamma^{kw/4-(j+1)/2}e^{b(j+1)\phi}e^{i k w/2\, \tilde x}.
\end{equation}
Now the strings in the cigar with integer/half-integer winding will correspond to strings in the U(1) direction with integer/half-integer winding. This means that we have two branes in the product model. One located in the U(1) space at $x_0=\theta_c-\theta_0$, in the $H_3^+$ directions this will be a brane with labels $(\theta_0,r)$ as we see from \eqref{eq:derivedboundarycond}. Secondly there must be a brane at $x_0=\theta_c-\theta_0+\pi$ and in the $H_3^+$ directions it will have label $(\theta_0,-r)$ or equivalently $(\theta_0+\pi,r)$. We can label the two branes $(\theta_c-\theta_0,(\theta_0,r))$ and $(\theta_c-\theta_0+\pi,(\theta_0,-r))$.

If we start with a string with zero winding in the cigar it will map to a string with zero winding in the U(1) direction and which starts and ends on the same brane say $(\theta_c-\theta_0,(\theta_0,r))$. If we do a half-integer spectral flow we get a string in the cigar going between two branches of the same brane, but in the product theory it will now be half winding in the U(1) direction and going between the two branes in the product theory. This is sketched in figure \ref{fig:h3branes}, see also figure 6 in \cite{RS}.

We can also relate the above discussion to the boundary action. We know what the boundary action looks like in the $H_3^+$ model for a single brane. The action in a first order formalism for the gauged theory must be exactly the same i.e.
 \begin{align}
 S_\text{bint} = i e^{i \theta_c}\lambda_B \int d u \beta e^{b \phi}\ ,
\end{align}
since this is nicely gauge invariant. However when we decouple $x$ in the rest of the action it will still appear in the boundary term:
\begin{align}\label{eq:brdryderived}
 S_\text{bint} = i \lambda_B e^{-i (x_0-\theta_c)} \int d u \beta e^{b \phi}.
\end{align}
This corresponds to eq. \eqref{eq:derivedboundarycond}.

\subsubsection*{Chan-Paton factors}
In order to calculate in the path integral we introduce Chan-Paton factors corresponding to the two branes. The new feature is that the boundary action will depend on which brane we are on, i.e. it will depend on the Chan-Paton factor. From \eqref{eq:brdryderived} we see that the form is exactly what we wanted, i.e.
\begin{align}\label{}
 S_\text{bint} = i \lambda_B e^{i \theta_0}\sigma_3 \int d u \beta e^{b \phi}.
\end{align}
Here have taken the brane $(\theta_c-\theta_0,(\theta_0,r))$ as the first brane and $(\theta_c-\theta_0+\pi,(\theta_0,-r))$ as the second. Boundary operators with half-integer winding in the U(1) direction should have a Chan-Paton factor that anti-commutes with this i.e. be $\sigma_1$ or $\sigma_2$. To discriminate between these two we note that the whole theory including branes is invariant under parity transformation in the $\beta\gamma$-system together with a $\pi$ rotation in the U(1) direction. This will map our two branes onto each other. The map of the fields to the cigar will be exactly the same under this conjugation. The conjugation is used in the main text to fix the Chan-Paton factors.

\section{Change of variables}
\label{Sklyanin}

When we relate the correlation functions of the cigar model to those of
Liouville theory plus a free boson, we utilize a change of variables
\begin{align}   \label{calB0}
&\sum_{a=1}^N \frac{\mu_a}{w - z_a}
 + \sum_{a=1}^N \frac{\bar \mu_a}{w - \bar z_a}
 + \sum_{b=1}^M \frac{\nu_b}{w - u_b}  \\
 & \qquad= u \frac{(w - \xi)^S \prod_{a' = 1}^{N '} (w - y_{a'}) (w - \bar y_{a'})
   \prod_{b' = 1}^{M '}  (w - t_{b'})  }
    {\prod_{a = 1}^{N } (w - z_{a}) (w - \bar z_{a})
   \prod_{b = 1}^{M }  (w - u_{b}) } \nonumber
\end{align}
as in \eqref{calB} and \eqref{calB2} subject to constraints \eqref{constmu}.
The Jacobian due to the change of
variables is used as in \eqref{jacobian} with \eqref{theta}, and we
would like to derive it in this appendix. For the bulk case,
the Jacobian due to the change of variables was
obtained in \cite{RT,Ribault} and extended to the case with
closed Riemann surfaces of arbitrary genus in \cite{HS3}.
For the boundary case, the variables are changed from
$(\mu_a , \bar \mu_a , \nu_b)$ to $(u, y_{a'}, \bar y_{a'} , t_{b'})$,
where there are $N'$ $y$'s and $M'$ $t$'s with
$M' + 2 N' + S +2 = M + 2 N$.
One non-trivial point here is that the number of $y_{a'}$ and $t_{b'}$
 changes when we vary $\mu_a,\nu_b$.
Thus we first show that the map \eqref{calB0} gives  a one-to-one map
modulo permutations among $y_{a'}$ and $t_{b'}$,
and hence the integral regions of the both side match.
Then we obtain the weight of the integral as given in the Jacobian
\eqref{jacobian}. For $S=0$ it was already given in \cite{HR}.

In order to show that \eqref{calB0} defines a
one-to-one map, we start with the map from $(u, y_{a'}, \bar y_{a'} , t_{b'})$
to $(\mu_a , \bar \mu_a , \nu_b)$. {}From the residues of
\eqref{calB0} at $z = z_a$ and $z = u_b $, we have
\begin{align}
 \mu_a &= u \frac{ (z_a - \xi)^S \prod_{a' = 1}^{N '} (z_a - y_{a'}) (z_a - \bar y_{a'})
   \prod_{b' = 1}^{M '}  (z_a - t_{b'})  }
    {(z_a - \bar z_a ) \prod_{c = 1, c \neq a}^{N }
 (z_a - z_c) (z_a - \bar z_c)
   \prod_{b = 1}^{M }  (z_a - u_{b}) } ~, \\
 \nu_b & =   u \frac{ (u_b - \xi)^S \prod_{a' = 1}^{N '}
 (u_b - y_{a'}) (u_b - \bar y_{a'})
   \prod_{b' = 1}^{M '}  (u_b - t_{b'})  }
    {\prod_{a = 1}^{N } | u_b - z_{a} |^2
   \prod_{d = 1, d \neq b}^{M }  (u_b - u_d) } ~,
\end{align}
and similarly for $\bar \mu_a$.
Therefore, if we choose the values of
$(u, y_{a'}, \bar y_{a'} , t_{b'})$, then we can obtain
$(\mu_a , \bar \mu_a , \nu_b)$ uniquely from the above equations.
A point here is that we can use arbitrary numbers of $y_{a'}$ and $t_{b'}$
if they satisfy $2 N ' + M'  = 2 N + M - S - 2 $.

To show that the map is onto the whole region of $(\mu_a , \bar \mu_a , \nu_b)$
we consider a given value of these and rewrite
\begin{align}
 &\sum_{a=1}^N \frac{\mu_a}{z - z_a}
 + \sum_{a=1}^N \frac{\bar \mu_a}{z - \bar z_a}
 + \sum_{b=1}^M \frac{\nu_b}{z - u_b} \\
  & \qquad = u \frac{(z - \xi)^S (
z^{2N'+M'} + a_1 z^{2N'+M'-1} + \cdots +  a_{2N'+M'} )}
    {\prod_{a = 1}^{N } (z - z_{a}) (z - \bar z_{a})
   \prod_{b = 1}^{M }  (z - u_{b}) } \nonumber
\end{align}
with $u= \sum_a 2 \, \text{Re}\, \mu_a z_a  + \sum_b \nu_b u_b$.
The term proportional to $z^{2N+M-1} = z^{2N' + M'+ S +1}$ vanishes due to the
delta function $\delta(\ell_0(\xi))$, where $\ell_n(\xi)$
is given by \eqref{constmu}. With the other delta functions
$\delta(\ell_n(\xi))$ with $n=1,2,\cdots,S$, the left hand side can be
factorized by $(z-\xi)^S$.
Since the left hand side is always real when we set $z = t$
with $t \in \mathbb{R}$, the coefficients $a_i$ have to be real.
Here we use the theorem that an algebraic equation of order $P$
with real valued
coefficients have $N'$ complex roots along with their complex conjugates
and $M'$ real roots with $2N' + M' = P$.
Therefore, once we choose some values of  $(\mu_a , \bar \mu_a , \nu_b)$,
then we can find $(u, y_{a'}, \bar y_{a'} , t_{b'})$ from the roots
of the above algebraic equation of order $2N' + M'$
with real coefficients up to permutations among $y_{a'}$ and $t_{b'}$.

In this way, we have shown that the map \eqref{calB0} is one-to-one between the
two regions of parameters. Therefore, in order to establish the
relation \eqref{jacobian} we just need to obtain the weight of
measure. In the case with $S=0$,
the Jacobian for the change of parameters was already
given in \cite{HR} as \eqref{jacobian} with $S=0$.
It can be also obtained by utilizing the bosonization formula for
$(b,c)$-ghosts as in appendix C of \cite{HS3}.
The correlation function we should use is
\begin{align}
 \left \langle \prod_{a=1}^N c(z_a) \bar c (\bar z_a )
 \prod_{i=1}^M c (u_i) \prod_{a'=1}^{N'}
 b(y_{a'}) \bar b(\bar y_{a'}) \prod_{i'=1}^{M'} b (t_{i'})
\right \rangle ~.
\end{align}
Notice that the point $t_{i'} = t_{j'}$ is really a singularity
since we cannot across the point continuously contrary to the
point $y_{a'} = y_{b'}$. Indeed, this is the
point where the number of $t_{i'}$'s would change. The same is true for
the point $y_{a'} = \bar y_{a'}$ where the number of $y_{a'}$'s would change.
The generalization to the case with $S \neq 0$ is actually quite
straightforward if we utilize the method in \cite{HS3}.
In that paper, the Jacobian is found by induction in $S$, and with the same trick
the Jacobian is found as \eqref{jacobian} with \eqref{theta}.

\section{Reflection relations of boundary operators}
\label{ra}

In the derivation of the FZZ duality, we utilize reflection relations of
bulk and boundary operators in the Liouville field theory with the action
\begin{align}
 S = \frac{1}{ \pi} \int d^2 w \left(
 \partial \phi \bar \partial \phi + \frac{\sqrt{g}}{4}
 {\cal R} Q \phi + \mu \pi e^{2 b \phi}  \right) + \mu_B \sigma_1
\int d u e^{b \phi }  ~.
\label{actionrr}
\end{align}
Here the parameters are  $Q=b+b^{-1}$, $b= i/\sqrt{2}$, $\mu = - 1$ and $\mu_B = - 1/\sqrt{2}$.
Notice that this Liouville field theory is different from the one obtained
from the relation to $H_3^+$ model. Rather, it is obtained by treating the
extra insertions as the Liouville interaction terms.
The conformal dimension of the bulk operator
$V_\alpha = \exp (2 \alpha \phi)$ is given by
$\Delta_\alpha = \alpha (Q - \alpha)$, which implies that
$V_\alpha$ is related to $V_{Q - \alpha}$ since they have the
same conformal dimension. In fact, we have the
reflection relation as $V_{\alpha}  = D(\alpha ) V_{Q -\alpha}$,
where $D(\alpha)$ can be obtained from the two point function.
With $b=i/\sqrt{2}$ the reflection coefficient is simplified as \cite{HS3}
\begin{align}
 D(\alpha ) = - (- \mu \pi)^{-1 + 2 \sqrt2 i \alpha}
  \frac{\Gamma (1 - \sqrt2 i \alpha)}
   {\Gamma ( \sqrt2 i \alpha )}
   \label{bulkrr}
\end{align}
with $\mu = - 1$.
In the same way, we should have reflection relations for the
boundary operators as
\begin{align}
[ B^i_\beta (z) ]_{s_1,s_2} = d (\beta| s_1, s_2) ^i_{~j}
 [ B^j_{Q - \beta} (z) ]_{s_1,s_2}  ~, \qquad
 B_\beta (z) = \frac{\sigma_i}{\sqrt{2}} e^{\beta \phi} ~.
 \label{boundaryrr}
\end{align}
The aim of this appendix is to compute the coefficient $d (\beta| s_1, s_2) ^i_{~j}$
with our specific values of parameters.
Since the boundary action includes the Chan-Paton factor,
the relations are different from those obtained
for the case without Chan-Paton factor \cite{FZZ,Teschner:2000md}.
Here we derive the reflection relations from
those without Chan-Paton factor. The idea is to change the
basis for the boundary operators such that the Chan-Paton factor in the boundary action is diagonal and
we can apply the formula for the FZZT-branes.

Before introducing the Chan-Paton factor, we summarize some useful
formulas for the FZZT-branes in \cite{FZZ,Teschner:2000md}.
The action of the boundary Liouville field theory is given as
\begin{align}
 S = \frac{1}{ \pi} \int d^2 w \left(
 \partial \phi \bar \partial \phi + \frac{\sqrt{g}}{4}
 {\cal R} Q \phi +  \pi \mu e^{2 b \phi}  \right)
 + \mu_B \int d u e^{b \phi } ~.
\end{align}
The boundary parameter $\mu_B$ may be expressed by $s$ as
\begin{align}
 \cosh \pi b s = \mu_B \sqrt{\frac{\sin \pi b^2 }{\mu}} ~.
 \label{paras}
\end{align}
The reflection relation for boundary action is then written as
\begin{align}
[ B_\beta (z) ]_{s_1,s_2} = d(\beta| s_1, s_2)
 [ B_{Q - \beta} (z) ]_{s_1,s_2}  ~, \qquad
 B_\beta (z) = e^{\beta \phi} ~. \label{boundaryfzz}
\end{align}
The reflection coefficient is found as
\begin{align}
 d(\beta|s_1 , s_2) = \frac{(\pi \mu \gamma (b^2) b^{2-2b^2})^{(Q-2\beta)/2b}
   \mathbf{G}(Q-2\beta) \mathbf{G}^{-1}(2\beta - Q)}
 {\mathbf{S} (\beta + \frac{i}{2} (s_1 + s_2)) \mathbf{S} (\beta - \frac{i}{2} (s_1 + s_2))
 \mathbf{S} (\beta + \frac{i}{2} (s_1 - s_2)) \mathbf{S} (\beta - \frac{i}{2} (s_1 - s_2))} ~,
\end{align}
where the functions $\mathbf{G}(x)$ and $\mathbf{S}(x)$ are defined as
\begin{align}
 &\log \mathbf{S} (x)
  = \int_0^{\infty} \frac{dt}{t}
 \left[ \frac{\sinh (Q - 2 x) t}{2 \sinh (bt) \sinh (t/b)}
 - \frac{(Q/2 - x)}{t} \right] ~, \label{S}\\
 &\log \mathbf{G}(x)
  =  \int_0^{\infty} \frac{dt}{t}
 \left[ \frac{e^{- Qt/2} - e^{-xt}}{(1 - e^{-bt}) (1 - e^{-t/b})}
  + \frac{(Q/2 - x)^2}{2} e^{-t}  + \frac{(Q/2 - x)}{t} \right] ~. \label{G}
\end{align}
We will use the shift relations
\begin{align}
 \mathbf{S} (x + b) = 2 \sin (\pi b x) \mathbf{S} (x) ~, \qquad
 \mathbf{S} (x + 1/b) = 2 \sin (\pi x/b) \mathbf{S} (x) ~.
 \label{shift}
\end{align}
For more details, see \cite{FZZ,Teschner:2000md}.

Now we include the effects of Chan-Paton factor.
The action we consider is \eqref{actionrr} and the boundary
operator is in \eqref{boundaryrr}. For the boundary operators,
it is convenient to take linear combinations as
\begin{align}
 B^{\pm}_\beta (u) = \frac{1}{2}(\sigma_0 \pm \sigma_1) e^{ \beta \phi} ~,
 \qquad
 B^{'\pm}_\beta (u) = \frac{1}{2}(\sigma_3 \mp i \sigma_2) e^{ \beta \phi} ~.
\end{align}
Notice that they behave as
\begin{align}
\sigma_1 B^{\pm}_\beta = \pm B^{\pm}_\beta ~, \qquad
B^{\pm}_\beta \sigma_1 = \pm B^{\pm}_\beta ~, \qquad
\sigma_1 B^{' \pm}_\beta = \pm B^{' \pm}_\beta ~, \qquad
B^{' \pm}_\beta \sigma_1= \mp B^{' \pm}_\beta , \label{behavior}
\end{align}
when the Chan-Paton factor in the boundary action acts on them.
Let us start with $B^+ _ \beta$. It is easy to see that the
reflection relation is the same as for the FZZT-branes.
Therefore, we
obtain $B^+ _ \beta = d(\beta|s,s) B^+ _ { Q -\beta }$,
where $s =i 5/4 b^{-1}$ from \eqref{paras} and the explicit
values of $\mu,\mu_B$.
Here, and in the following, we restrict the domain of $s$ as
$0 \leq - i b s < 2$ to avoid ambiguity.
For $B^- _ \beta$, we may shift $s \to s - i/b$
since $\sigma_1 B^-_\beta = B^-_\beta \sigma_1
= - B^{-}_\beta $ with \eqref{paras}.
Thus the reflection coefficient is
\begin{align}
d (\beta | s - \tfrac{i}{b} , s - \tfrac{i}{b})  =
 \frac{\sin (\frac{\pi}{b}
  (\beta + i s - \frac{1}{b}))}{\sin (\frac{\pi}{b}
  (\beta - i s ))}
   d (\beta | s , s )= - d (\beta | s , s )
\end{align}
where we have used the shift relations \eqref{shift}.
Moreover, we have set $1/b^2 = -2$ and $s =i 5/4 b^{-1}$.
The reflection relations are thus
\begin{align}
B^0_{\beta} = d(\beta | s , s )
B^1_{Q - \beta} ~, \qquad
B^1_{\beta} = d(\beta | s , s )
B^0_{Q - \beta}
\end{align}
in the original basis.
We now move to the other cases with $B^{'\pm}_\beta$.
{}From \eqref{behavior}, we propose that the reflection relations
are given as%
\begin{align}
B^{'+}_{\beta}
 = d(\beta | s  - \tfrac{i}{2b}, s + \tfrac{i}{2b} ) B^{'+}_{Q - \beta} ~,
 \qquad
B^{'-}_{\beta}
 = d(\beta | s  + \tfrac{i}{2b}, s - \tfrac{i}{2b})  B^{'-}_{Q - \beta} ~,
\end{align}
where we have used $s =i 5/4 b^{-1}$. There might be other choices,
but it turns out that they do not work well.
Noticing that $b^{-1} = - 2 b$, we can show
\begin{align}
d(\beta |s  - \tfrac{i}{2b}, s + \tfrac{i}{2b} )
 = \cot ( \pi b \beta ) d(\beta | s , s )
= d(\beta |s + \tfrac{i}{2b}, s - \tfrac{i}{2b} ) ~.
\end{align}
Therefore, we have
\begin{align}
 B^2_{\beta}
  =  \cot ( \pi b \beta ) d(\beta | s , s )
  B^2_{Q-\beta} ~, \qquad
 B^3_{\beta}
  = \cot ( \pi b \beta ) d(\beta | s , s )
  B^3_{Q-\beta}
\end{align}
in the original basis.

Above, we have seen that the reflection coefficient in \eqref{boundaryrr} can be
written in terms of $d (\beta| s, s) $ for the FZZT-branes defined in \eqref{boundaryfzz}.
In fact, we can show that  the function
$d(\beta|s,s)$ with $b=i/\sqrt{2}$ and $s =  i 5/4 b^{-1}$
has a simple expression.
Since it is quite difficult to directly compute the
functions \eqref{S} and \eqref{G}, we take a different route.
We choose to utilize the relation shown in \cite{FZZ}
\begin{equation}\label{eq:relation2pt}
d(\beta|s,s) \ = \ c_-(\beta) d(\beta+b|s,s) ~,
\end{equation}
which the reflection coefficient has to satisfy.
Here $c_-(\beta)$ is simplified as
\begin{equation}
\begin{split}
c_-(\beta) \ &= \ ( - 4 \mu /\pi)\sin^2(\pi\beta b)
 \sin(\pi\beta b+\pi/4)\sin(\pi\beta b-\pi/4) \\  & \qquad \qquad \times
\Gamma(1-2\beta b)\Gamma(2\beta b -1) \Gamma(3/2-2\beta b)\Gamma(2\beta b-1/2)\\
&= \ - \pi\mu \frac{\sin(\pi\beta b)}{\cos(\pi\beta b)}\frac{\Gamma(2(\beta+b) b)}{\Gamma(2\beta b )}
\end{split}
\end{equation}
for our choice of $s$.
The relation \eqref{eq:relation2pt} can be solved quite easily.
Assigning the unitarity relation
\begin{equation}
d(\beta)d(Q-\beta)  =  1 ~,
\end{equation}
we find for the reflection coefficient
\begin{equation}
d(\beta|s,s) = (- \mu \pi)^{(Q - 2\beta) /(2b)} \sqrt{\frac{\pi}{2}}
 \frac{1}{\Gamma(2\beta b)
\sin (\pi b \beta + \pi /2 )}  ~.
\end{equation}
We also need
\begin{align}
  \cot (\pi b \beta) d(\beta|s,s) =
(- \mu \pi)^{(Q - 2\beta) /(2b)} \sqrt{\frac{\pi}{2}}
 \frac{1}{\Gamma(2\beta b)
\sin (\pi b \beta )} ~.
\end{align}

\section{Branes in sine-Liouville theory}
\label{SLbranes}

Branes in sine-Liouville theory have not been investigated yet.
However, it is easy to guess what they are by two means.
Firstly, one can utilize the boundary FZZ duality, but
of course we cannot use this to prove the duality itself.
Secondly, one can mimic the arguments in \cite{Hosomichi} for
branes in ${\cal N}=2$ Liouville theory.
The details may be changed, but qualitatively
they should be the same as branes in sine-Liouville theory.
In particular, we should be able to obtain the branes in sine-Liouville theory by
repeating the analysis in \cite{Hosomichi}.

Following the terminology in \cite{Hosomichi}, we call D1-branes for A-branes
and D2-branes for B-branes. A-branes are relatively easy since we can
construct boundary states for them only from modular transformations
of annulus amplitudes. According to \cite{Hosomichi}, we cannot
construct B-branes from modular invariance of annulus amplitudes,
and we have to study the factorization constraint from the
two-point function on a disk.
Fortunately, we have shown that B-branes in sine-Liouville
theory can be obtained from D1-branes in the cigar model, so we just need
to study A-branes.

Since the branes are labeled by the representation of $\text{sl}(2,\mathbb{R})$,
A-branes may be classified according the representation as in section 3.1
of \cite{Hosomichi}. Following the paper, we have six types as
\begin{enumerate}
\item Identity representation
\item Non-chiral non-degenerate representations
\item Non-chiral degenerate representations
\item Anti-chiral representations
\item Chiral representations
\item Degenerate representations
\end{enumerate}
Types 2,4,5 and types 1,6 would correspond to D2-branes and
D1-branes in the cigar model, respectively.
Type 3 requires some consideration.
We just need to repeat the analysis in \cite{Hosomichi} to construct
boundary states for the above branes, but replacing the characters
for the ${\cal N}=2$ superconformal algebra by those for the $\hat W_\infty (k)$-algebra
(or equivalently parafermions), see section 4 of \cite{Bakas:1991fs}.

In order to go further, we need to compute two-point functions
of bulk operators on a disk with one degenerate operator and
two-point functions of boundary operators on a disk.
Even though it is a very hard task to compute them directly,
it is easy to guess the boundary actions.
For B-branes we have already obtained this as
\begin{align}
 {\cal L}_\text{bdy}
 =  \left( \mu_B \sigma_+
 + \mu_{\bar B} \sigma_-  \right)
 e^{\frac{1}{2 b}\phi + i \frac{\sqrt{k}}{2} \tilde X  }
+
 \left(\bar \mu_B \sigma_+ + \bar \mu_{\bar B} \sigma_- \right)
 e^{\frac{1}{2 b}\phi - i \frac{\sqrt{k}}{2} \tilde X  }
~, \nonumber
\end{align}
with
\begin{align}
 (\mu_{B}, \mu_{\bar B}, \bar \mu_{\bar B}, \bar \mu_{\bar B})
  = i \sqrt{\frac{\tilde c}{2}}
   (e^{i \pi (J - M)} , e^{- i \pi (J - M)} ,e^{- i \pi (J + M)} ,e^{i \pi (J + M)} ) ~.
\end{align}
This form of the action implies that B-branes consist of two D2-branes
and a non-trivial Wilson loop is included.
For A-branes corresponding to D2-branes in the cigar model, we
propose the following boundary action (see (5.28) in \cite{Hosomichi})
\begin{align}
 {\cal L}_\text{bdy}
 = \mu_A \sigma_+
 e^{\frac{1}{2 b}\phi + i \frac{\sqrt{k}}{2} \tilde X  }
+ \bar \mu_A \sigma_-
 e^{\frac{1}{2 b}\phi - i \frac{\sqrt{k}}{2} \tilde X  }
\end{align}
with
\begin{align}
 \mu_A = \sqrt{2 \tilde c} \sin \pi (J - M) ~, \qquad
 \bar  \mu_A = \sqrt{2 \tilde c} \sin \pi (J + M) ~.
\end{align}
This form of the action implies that A-branes consist of two D1-branes
and tachyonic states from open strings between two branes are condensed.
Following the analysis in \cite{FL} we can show that the boundary action
preserves the $\hat W_\infty (k)$ symmetry of sine-Liouville theory,
just like the boundary action for
A-branes in the ${\cal N}=2$ super Liouville theory preserves superconformal symmetry. Therefore, the problem left is to fix the constant
coefficients including the Chan-Paton factors.
However, in order to show that our proposal is correct, we would need to study
correlation functions.

\end{document}